\documentclass{article}
\usepackage{amsfonts}
\usepackage{amssymb}
\usepackage{amsmath}
\usepackage{graphicx}
\usepackage{slashed}
\usepackage{setspace}
\usepackage{simplewick}
\usepackage{cite}

\def\half{{1\over 2}}

\newcommand{\reef}[1]{(\ref{#1})}
\newcommand{\rom}[1]{\mathrm{#1}}
\newcommand{\example}{\noindent\hangindent 1cm\hangafter0{\bf Example:}~~}

\newcommand{\A}{{\cal A}}

\newcommand{\MHV}{\A^{\rm MHV}}
\newcommand{\cyc}{{\rm cyc}}

\newcommand{\cn}{{\cal N}}

\newcommand{\cf}{{\cal F}}

\newcommand{\be}{\begin{equation}}
\newcommand{\ee}{\end{equation}}

\def\be{\begin{equation}}
\def\ee{\end{equation}}
\def\bea{\begin{eqnarray}}
\def\eea{\end{eqnarray}}
\def\ba{\begin{array}}
\def\ea{\end{array}}
\def\bd{\begin{displaymath}}
\def\ed{\end{displaymath}}

\def\ie{{\it i.e.~}}

\def\a{\alpha}
\def\b{\beta}

\def\d{\delta}
\def\e{\epsilon}

\def\g{\gamma}
\def\h{\eta}

\def\l{\lambda}
\def\m{\mu}

\def\pa{\partial}

\def\>{\rangle}
\def\<{\langle}
\def\Dsl{D \hskip-.6em \raise1pt\hbox{$ / $ } }
\def\to{\rightarrow}

\def\pa{\partial}

\def\lab{\label}

\newcommand{\eps}{\epsilon}
\newcommand{\lra}{\leftrightarrow}

\def\da{{\dot\alpha}}

\def\tQ{\tilde{Q}}

\begin{document}

\setstretch{1.05}

\begin{titlepage}

\begin{flushright}
MIT-CTP-4000 \\
UUITP-26/08
\end{flushright}
\vspace{1cm}

\begin{center}
{\Large\bf Proof of the MHV vertex expansion } \\[2.5mm]
{\Large\bf for all tree amplitudes in $\mathbf{\cn =4}$ SYM theory }\\
\vspace{1cm}
{\bf Henriette Elvang${}^{a}$\footnote{On leave of absence from Uppsala University.},
Daniel Z.~Freedman${}^{b,c}$, Michael Kiermaier$^{b}$} \\
\vspace{0.7cm}
{{${}^{a}${\it School of Natural Sciences}\\
         {\it Institute for Advanced Study}\\
         {\it Princeton, NJ 08540, USA}}\\[5mm]
{${}^{b}${\it Center for Theoretical Physics}}\\
{${}^{c}${\it Department of Mathematics}}\\
         {\it Massachusetts Institute of Technology}\\
         {\it 77 Massachusetts Avenue}\\
         {\it Cambridge, MA 02139, USA}}\\[5mm]
{\small \tt  elvang@ias.edu,
 dzf@math.mit.edu, mkiermai@mit.edu}
\end{center}
\vskip .3truecm

\begin{abstract}
\noindent

We prove the MHV vertex expansion for all tree amplitudes of $\cn =4$  SYM theory. The
 proof uses a shift acting on all external momenta, and  we show that every N$^k$MHV tree
 amplitude falls off as $1/z^k$, or  faster, for large $z$ under this shift. The MHV vertex
 expansion allows  us to
  derive compact and efficient
 generating functions for all N$^k$MHV tree
 amplitudes of the theory.  We
 also
derive an improved form of the anti-NMHV generating function. The proof leads to a curious set of sum
rules for the diagrams of the MHV vertex expansion.

\end{abstract}

\end{titlepage}

\setstretch{0.5}
\tableofcontents
\setstretch{1.05}

\newpage
\setcounter{equation}{0}
\section{Introduction}

There has been remarkable progress in calculations  of on-shell tree  amplitudes since the advent of the modern form of recursion relations \cite{csw,bcf,bcfw}, with
 many useful applications in QCD, $\cn =4$ super Yang-Mills theory (SYM), general relativity, and $\cn =8$ supergravity (see, for instance, \cite{bdkreview} and references therein).
Through generalized unitarity cuts \cite{bddk,Bern:1994cg,Bern:1996ja,Bern:1997nh,Bern:2004ky,Bern:2005iz,Bern:2007ct,Cachazo:2008dx,Forde:2007mi,Bern:2007hh},  tree amplitudes play a central role as building blocks for loop amplitudes. It is therefore important to have reliable and efficient methods to calculate them.
The purpose of this paper is to achieve this goal for all tree amplitudes of 4-dimensional $\cn=4$ SYM theory.

On-shell $n$-point amplitudes $\A_n$ of $\cn=4$ SYM can be
classified within sectors denoted by N$^k$MHV.  For fixed $k$ and $n$ each sector contains the amplitudes with $k+2$ negative and $n-k-2$ positive helicity gluons together with all other amplitudes related to these by SUSY Ward identities \cite{Grisaru:1976vm,gris,bdp}.
MHV amplitudes are particularly simple since SUSY fixes them completely in terms of the MHV gluon amplitude $\A_n(- -+\dots+)$, which is given by the Parke-Taylor formula \cite{pt}.

Recursion relations are powerful because they express $\A_n$ in terms of lower-point on-shell amplitudes, and this allows a recursive construction which is much more efficient than Feynman diagram calculations.  We are particularly interested in the type of recursion relation which leads to the \emph{MHV vertex expansion}, also known as the  \emph{CSW expansion} \cite{csw}. Here every lower-point amplitude is MHV and that makes the computation of amplitudes with general external states very efficient.

Our paper has three main results. The first is the proof that the MHV vertex expansion is valid for all N$^k$MHV tree amplitudes of $\cn=4$ SYM theory. The expansion contains a sum of  diagrams,  each with $k+1$ vertices  connected by $k$ internal propagators. Each vertex is an on-shell MHV subamplitude. The MHV vertex expansion provides a simple diagrammatic method for explicit calculation of tree amplitudes for any set of external states of the $\cn=4$ theory.

The second main result is a systematic construction of \emph{generating functions} $\cf^\text{N$^k$MHV}_n$
 which package the MHV vertex expansion of all N$^k$MHV $n$-point amplitudes. The  generating functions $\cf^\text{N$^k$MHV}_n$  depend on Grassmann-valued bookkeeping variables, and the states of $\cn =4$ SYM are in $1:1$ correspondence with Grassmann differential operators involving these variables \cite{BEF}. The MHV vertex expansion of any desired amplitude is then obtained  by applying the differential operators associated with the external states to $\cf^\text{N$^k$MHV}_n$.  This builds on, and extends, the work \cite{nair,ggk}.
The process of computing specific amplitudes is easily automated in combined symbolic and numerical computer codes. We use these codes in extensive checks of our construction.

Thirdly, anti-N$^k$MHV generating functions can be obtained from N$^k$MHV generating functions by a simple prescription. At the anti-NMHV level, we present a new simple form of the generating function. In the process of deriving it, we find new sum rules for MHV vertex diagrams.

Several issues motivate our work.  One is that the validity of recursion relations is sensitive to the spins of the external particles. For instance  there are amplitudes in $\cn = 8$ supergravity for which the standard MHV vertex method fails \cite{BEF}.
Thus, although the MHV vertex expansion is proven for pure Yang-Mills theory \cite{risager}, it cannot be assumed without proof that it is valid for all amplitudes of $\cn = 4$ SYM.
An interesting point is that the MHV vertex expansion is simpler for amplitudes with a generic set of external states of $\cn=4$ SYM: fewer diagrams are non-vanishing than for amplitudes with external gluons only.

In \cite{BEF} and \cite{efk1} it was shown in many examples how the use of generating functions for MHV and NMHV amplitudes simplifies the intermediate state helicity sums which are an essential
part of the unitarity method used to obtain loop amplitudes from products of trees.
Furthermore, using the KLT relations \cite{Kawai:1985xq} and the map between $\cn=8$ supergravity and the
direct product of two copies of $\cn=4$ SYM,  helicity sums in the supergravity theory can be computed via helicity sums in the $\cn=4$ theory.  One must
anticipate that multi-loop investigations in both $\cn=4$ and $\cn =8$ will employ the MHV vertex expansion at higher N$^k$MHV levels, so it is vital  that the expansion is rigorously valid.

Recursion relations
can be derived from the analyticity and pole factorization of tree amplitudes in a complex variable $z$ associated with a deformation or shift of a subset of the external momenta.  If the shifted  amplitude vanishes as $z \to \infty$, Cauchy's theorem can be used to establish a valid recursion relation for that amplitude. Amplitudes and subamplitudes are written in the spinor-helicity formalism in which complexified null 4-momenta $p_i^\m$ are described using spinors
$\l_i^{\dot{\a}} \lra |i\>$ and $\l_i^\a \lra |i].$

BCFW recursion relations \cite{bcfw} are obtained from a shift of the spinors $|i],~ |j\>$ of two external momenta
$p^\m_i,~p^\m_j$ which can be chosen arbitrarily. Following \cite{nima1}, it was recently shown \cite{ccg} that any
$n$-point amplitude in which at least one particle is a negative helicity gluon (and the remaining particles
are any set of vectors, spinors, or scalars) admits a BCFW shift under which the amplitude
vanishes for large $z$.  In \cite{efk1} we used SUSY Ward identities to extend this result and prove that \emph{all amplitudes of $\cn = 4$ SYM admit valid BCFW recursion relations.}\footnote{With one exception: the MHV 4-scalar amplitude $\big\< A^{12} \, A^{23}\, A^{34}\, A^{14} \big\> = 1$ has no poles, so no factorization is possible. In a Feynman diagram calculation, only the 4-point contact term contributes to the amplitude.}
In the present paper we use this extension to obtain a valid BCFW representation of all N$^k$MHV amplitudes with $k \ge 1$. This representation allows us to prove that the amplitudes vanish under an all-line shift (described below), whose associated recursion relations eventually lead to the MHV vertex expansion.

The MHV vertex expansion for $n$-gluon N$^k$MHV amplitudes was proven in \cite{risager} using a shift  of the square spinors $|i]$ of the $k+2$ negative helicity external lines.
In \cite{efk1}  we used a 3-line shift in which the particles on the shifted lines carry at least one common $SU(4)$ index to validate the MHV vertex  expansion for all NMHV amplitudes in the $\cn =4$ theory.
Similarly, a $(k+2)$-line common-index shift can be used to extend this result to all N$^k$MHV amplitudes, but the proof of large $z$ falloff turns out to be much simpler if we use
an \emph{all-line shift}, a shift of all spinors $|i], ~i=1,\ldots n$.
To reiterate: we have proven that both the common-index shift and the all-line shifts give the needed large $z$ falloff and that they both
lead to the MHV vertex expansion. For simplicity, we only present the proof involving the all-line shift.

To prove that  N$^k$MHV amplitudes vanish for large $z$ under the all-line shift we use the following strategy. We start with the valid BCFW recursion relation for any amplitude $\mathcal{A}_n^\text{N$^k$MHV}$ and study the effect of an all-line shift on each diagram of that recursion relation.  We show, using induction on $k$ and $n$, that each diagram of the analytically continued amplitude $\mathcal{A}^\text{N$^k$MHV}_n(z)$ vanishes at least as fast as $1/z^k$ as $z \to \infty.$
This is (more than) enough to establish the associated \emph{all-line recursion relation} for the amplitude. It expresses any N$^k$MHV amplitude in terms of N$^q$MHV subamplitudes with $q<k$. We then proceed,  again by induction,  and assume that the MHV vertex
expansion is valid at level N$^q$MHV and substitute that expansion into the subamplitudes. This substitution does not immediately give the MHV vertex expansion for the N$^k$MHV amplitude because the subamplitudes are evaluated at shifted momenta. However, we show (generalizing the approach of \cite{greatdane}) that the shift dependence cancels after combining sets of $2k$ diagrams, and this way we obtain the desired MHV vertex expansion.

{}From the MHV vertex expansion we derive the generating functions $\cf_n^\text{N$^k$MHV}$
from which any amplitude can be computed efficiently by applying the appropriate Grassmann derivative.
As the starting point of this derivation we use the well known MHV generating functions of Nair \cite{nair}.

In the original CSW proposal \cite{csw}, an arbitrary reference spinor $|X]$ was introduced in order to evaluate the subamplitudes of the MHV vertex expansion as on-shell expressions. The
individual diagrams of the MHV vertex expansion typically
depend on $|X]$, but it was argued that the full amplitude
is independent of $|X]$.  In the recursion relation approach, the reference spinor $|X]$ arises from the particular form of the shift,
for both the common-index shift and all-line shift, and again each diagram of the expansion
depends on $|X]$.  However, when the amplitudes vanish as $z \to \infty$ for all choices of $|X]$,  Cauchy's theorem ensures that the sum of contributing diagrams is $|X]$-independent.
The $|X]$-independence is a useful practical test of our arguments concerning large $z$ behavior and the construction of  generating functions.  We have
carried out numerical checks of $|X]$-independence in various examples.

An alternative to the MHV vertex method is the approach of
\cite{Drummond:2008cr}
in which generating functions, there called superamplitudes, are
expressed in terms of invariants of dual superconformal symmetry
\cite{Drummond:2006rz,Alday:2007hr,dks,Drummond:2007cf,Drummond:2007au,dhks,Alday:2007he,Brandhuber:2008pf,Beisert:2008iq,McGreevy:2008zy,Berkovits:2008ic,sok2}. The superamplitudes are obtained from an intriguing  form of recursion relations based on supershifts [\citen{nima2},\citen{Brandhuber:2008pf}].
We will comment briefly on these new developments in section \ref{s:disc}.

The logical organization of our argument proceeds from proving the large $z$
falloff needed by the all-line shift recursion relations, then from these recursion relations to the MHV vertex expansion, and finally to the expansion for generating functions. Rather than starting out with the most technical parts of the argument, we prefer to first introduce the MHV vertex expansion, then the generating functions and finally the proof of their validity.
Thus the outline of the paper is as follows:
In~section \ref{secnotation} we introduce notation and review the known structure of MHV amplitudes which are the building blocks of the MHV vertex expansion. We also discuss the MHV generating function, the map between particle states of $\cn=4$ SYM and Grassmann differential operators, and the supercharges in this formalism. In section \ref{secMHVVE} we discuss the structure of MHV vertex expansions and the diagrams they contain.
In section \ref{secgenfct}  we convert the expansions for individual amplitudes to generating functions $\cf_n^\text{N$^k$MHV}$.
In section \ref{secVEfromCIS},  we show that the MHV vertex expansion can be derived from  all-line shift recursion relations.
In section \ref{secvalidity}, we establish the large $z$ behavior required for all-line shift recursion relations.
In section \ref{secsquare}, we discuss the behavior of  MHV vertex diagrams under various square spinor shifts.
In section \ref{secanti}, we discuss generating functions for anti-N$^k$MHV amplitudes and derive new sum rules.
Section \ref{s:disc} is devoted to discussion of our work and other approaches.
Appendices \ref{appdeltaID} and \ref{app:shifts} contain material that belongs in appendices.


\section{Notation and review}\label{secnotation}

We use the spinor helicity formalism formalism with conventions given in Appendix A of \cite{BEF}.

The boson and fermions of $\cn =4 $ SYM theory are described by annihilation operators which are listed in order of decreasing helicity as
\be \lab{bofe}
B_+(i) \hspace{1cm} F_+^a(i)\hspace{1cm} B^{ab}(i) = \half \e^{abcd}B_{cd}(i)\hspace{1cm} F_a^-(i) \hspace{1cm} B^-(i)\, .
\ee
The argument $i$ indicates the particle momentum $p^\m_i$. The six scalars satisfy the displayed
$SU(4)$ self-duality condition. Under the global symmetry group $SU(4)$, particles transform in representations which are
anti-symmetric tensor products of the $4$ or $\bar{4}$, and
particles of opposite helicity transform in conjugate representations.
It is convenient to  ``dualize"
lower indices and use a notation in which the annihilation operators for all
particles carry upper indices only.  The list above is then replaced by
\bea \lab{annih}
\begin{split}
A(i) =B_+(i)\hspace{1cm} A^a(i)=F^a_+(i) \hspace{1cm} A^{ab}(i)=B^{ab}(i)\\[2mm]
 A^{abc}(i) =\e^{abcd}F^-_d(i)\hspace{1cm} A^{abcd}(i)=\e^{abcd} B^-(i)\, .
 \end{split}
\eea
Note that the helicity $h$ (and the bose-fermi statistics) is determined by the
number $r$ of indices carried by  operator $A^{a\dots }(i)$.  Indeed,
 $2h = 2 - r$.

Chiral supercharges
 $Q^a \equiv - \e^\a\,Q^a_\a$ and
$\tilde{Q}_a \equiv \tilde{\e}_{\dot{\a}} \tilde{Q}^{\dot{\a}}_a$ are defined to
include contraction with the  anti-commuting parameters $\e^\a, \tilde{\e}_{\dot{\a}}$ of SUSY transformations.
The commutators of the operators $Q^a$
and $\tilde{Q}_a$ with the annihilators are given by:
\bea \lab{comalg}
\begin{array}{rcl}
\big[\tQ_a,A(i)\big] &=& 0\, ,\\[2mm]
\big[\tQ_a,A^b(i)\big] &=& \<\epsilon\, i\>\,\d^b_a\,A(i)\, ,\\[2mm]
\big[\tQ_a,A^{bc}(p)\big]
&=&\<\epsilon\,i\> \, 2! \, \d^{[b}_a\,A^{\raisebox{0.6mm}{\scriptsize$c]$}}(i)\, ,\\[2mm]
\big[\tQ_a,A^{bcd}(i)\big] &=&\<\epsilon\, i\>\, 3!\, \d_a^{[b} A^{\raisebox{0.6mm}{\scriptsize$cd]$}}(i)\, ,\\[2mm]
\big[\tQ_a,A^{bcde}(i)\big] &=& \<\epsilon\,i\>\, 4!\, \d_a^{[b} A^{\raisebox{0.6mm}{\scriptsize$cde]$}}(i) \, ,
\end{array}
\hspace{8mm}
\begin{array}{rcl}
[Q^a,A(i)] &=& [i\,\epsilon]\,A^a(i)\, , \\[2mm]
\big[Q^a,A^b(i)\big] &=& [i\,\epsilon]\, A^{ab}(i)\, , \\[2mm]
\big[Q^a,A^{bc}(i)\big]
&=&[i\,\epsilon]\, A^{abc}(i)\, , \\[2mm]
\big[Q^a,A^{bdc}(i)\big]
&=& [i\,\epsilon]\,A^{abdc}(i)\, , \\[2mm]
\big[Q^a,A^{bcde}(i)\big] &=& 0\, .
\end{array}
\lab{n4tQ}
\eea
Note that $\tQ_a$  raises the helicity of all operators and involves
the spinor angle bracket $\<\e\,i\>$.  Similarly, $Q^a$
lowers the helicity and spinor square brackets $[i\, \e]$ appear.

We focus on annihilation operators because particles are treated as outgoing.
An $n$-particle amplitude can then be designated as
\be  \lab{genamp}
\A_n(1,\ldots,n)=\<A^{a_1 \ldots}(1) A^{a_2\dots}(2)\ldots A^{a_n\ldots}(n)\>\,,
\ee
and viewed as the indicated product of annihilators acting to the left on the ``out vacuum" state.  We deal exclusively with color-ordered tree amplitudes.

Amplitudes must be $SU(4)$ invariant.  An amplitude vanishes unless each index value
$a=1,2,3,4$ appears exactly $k+2$ times in the operators in \reef{genamp}.  An amplitude
then carries a total of $4k+8$  indices.   Amplitudes with a fixed value of $k$ comprise the N$^k$MHV sector discussed in the Introduction.

A special case is $k=-1$. SUSY Ward identities can be used to show that amplitudes with 4 indices vanish, unless they have just 3 external legs (and are evaluated with complex momenta). The non-vanishing N$^{-1}$MHV$_3$ = anti-MHV$_3$ amplitudes play an important role in the BCFW recursion relations we use in section \ref{secvalidity}.

MHV amplitudes ($k=0$) are the building blocks of the MHV vertex
expansion.  They include the $n$-gluon Parke-Taylor amplitude with two negative helicity gluons of momenta $p_i,\,p_j$
\bea   \lab{pt}
\bigl\<A(1)\ldots A^{1234}(i)\ldots A^{1234}(j) \ldots A(n)\bigr\>  &=& \frac{\<i\,j\>^4}{\cyc(1,\dots,n)}\, ,\\
\lab{cyc}
\cyc(1,\dots,n) &\equiv&  \prod_{i=1}^n \<i\,(i+1)\>\, \, ,
\eea
with the cyclic identification $|n+1\> = |1\>$.

The other amplitudes in the MHV sector are encoded in the generating function \cite{nair}
\be   \lab{mhvgen}
\cf_{n}^{\rm MHV} \,= \,  \frac {\d^{(8)}\bigl(\sum_{i=1}^n |i\> \h_{ia}\bigr)}{ \cyc(1,\dots,n) } \, .
\ee
The Grassmann $\d$-function is a sum of terms, each of which is a product of
8 bookkeeping variables
$\eta_{ia}$.
This is clear from the identity
\be \lab{dident}
\d^{(8)}\biggl(\,\sum_{i=1}^n |i\> \h_{ia}\biggr) = \frac{1}{16} \prod_{a=1}^4 \sum_{i,j=1}^n \<i\,j\> \h_{ia}\h_{ja}\,.
\ee
In \cite{BEF}, it was shown that the annihilation operators of \reef{annih} are associated with differential operators
\begin{equation}
\begin{split}
\lab{diffop}
A(i) &\lra 1 \hspace{0.6cm}   A^a(i) \lra D^a_i = \frac{\pa}{\pa \h_{ia}}  \hspace{0.8cm} A^{ab}(i) \lra D_i^{ab}= \frac{\pa^2}{\pa \h_{ia} \pa \h_{ib}} \\
A^{abc}(i) &\lra D_i^{abc} = \frac{\pa^3}{\pa \h_{ia}\pa \h_{ib}\pa\h_{ic}} \hspace{0.8cm}  A^{1234}\lra D^{(4)}_i = \frac{\pa^4}{\pa \h_{i1}\pa \h_{i2}\pa\h_{i3}\h_{i4}}\,.
\end{split}
\end{equation}
Any desired MHV amplitude is then obtained by applying the appropriate product of these operators, of total order 8, to the generating function.

\example Consider the 5-point function $\< A^1(1)A^{23}(2)A(3)A^{234}(4)A^{14}(5)\>$. Using the dictionary~(\ref{diffop}) to determine the differential operators that correspond to the external states of the amplitude, we obtain
\begin{equation}
\bigl\< A^1(1)A^{23}(2)A(3)A^{234}(4)A^{14}(5)\bigr\>~=~ D^1_1 D^{23}_2 D^{234}_4 D^{14}_5\; \cf_5^{\rm MHV}
~=~
 \frac{\<1\,5\>\<2\,4\>^2\<4\,5\>}{\cyc(1,\ldots,5)}\,.
\end{equation}

\noindent
Any MHV $n$-point function thus takes the form
\be \lab{genmhv}
\frac{\<\;\>\<\;\>\<\;\>\<\;\>}{\cyc(1,\dots,n)} \, .
\ee
The numerator is always a product of four spinor products $\<\;\>$ which we call the spin factor. It is obtained by applying the appropriate 8th order differential operator to the $\d$-function in \reef{dident}.   The calculation of spin factors
reduces to a simple   Wick contraction algorithm of the
 basic operators $\pa/\pa\h_{ia}$.
The elementary contraction is
\be \lab{wick}
\contraction[1.5ex]{}{\frac{\pa}{\pa\h_{ia}}}{~\ldots~}{\frac{\pa}{\pa\h_{jb}}}{}
\frac{\pa}{\pa\h_{ia}}~\ldots~\frac{\pa}{\pa\h_{jb}}
~=~ \pm \,\d^{ab}\<i\,j\>\,.
\ee
The $\ldots$ indicate other operators between the pair which is contracted, and
the sign depends on the number of these operators.

In \cite{BEF}  supercharges
\be \lab{schg}
|\tQ_a\> \,=\, \sum_{i=1}^n | i\>\, \h_{ia} \, ,
\hspace{1.5cm} [Q^{a}|  \,=\, \sum_{i=1}^n\, [i|\, \frac{\pa}{\pa \h_{ia}}
\ee
were defined. They act by multiplication and differentiation on functions of the $\h_{ai}$.   The SUSY algebra is obeyed, and the commutators of $\< \eps \, \tQ_a \>$ with
the differential operators of \reef{diffop} are isomorphic to the field theory relations \reef{comalg}. It is immediate  that $\< \eps\, \tQ_a\> \, \cf^{\rm MHV}_n=0$, while
$[Q^a  \,\eps] \, \cf^{\rm MHV}_n=0$ follows from momentum conservation.
As a consequence, amplitudes obtained from the generating function by the procedure we have outlined obey all SUSY Ward identities.


\setcounter{equation}{0}
\section{The MHV vertex expansion}\label{secMHVVE}
In this section we illustrate the structure of
 the MHV vertex expansion by studying the diagrams which contribute to
 N$^k$MHV amplitudes for the simplest cases $k=1,2,3$. This study motivates
 the formula for general $k$ given in  (\ref{AnNkMHV0}).
Our goal in this section
is a clear presentation of the ingredients of the expansion, rather than a derivation.
 The derivation will be presented in section~\ref{secVEfromCIS}.

\subsection{Structure of the MHV vertex expansion}\label{sec2diagrams}
Let us examine the MHV vertex diagrams which contribute to N$^k$MHV amplitudes for low $k$.  The simplest case  $k=1$ corresponds to  NMHV amplitudes. There is just one type of diagram consisting of
two MHV subamplitudes, called $I_1$ and $I_2$, connected by a propagator for an internal line $P_\a$.
Thus, for an $n$-point NMHV amplitude, the MHV vertex expansion gives
\begin{equation}\label{ANMHV}
\begin{split}
    \A^{\rm NMHV}_n(1,\ldots,n)
    &=\hskip-.3cm\sum_{\stackrel{\text{MHV vertex diagrams }}{\alpha}}\hskip-.5cm
    \frac{\MHV \!(I_1)\MHV \!(I_2)}{P_{\alpha}^2}
    =\hskip-.3cm\sum_{\stackrel{\text{MHV vertex diagrams }}{\alpha}}\hskip-.2cm
   \parbox[c]{4.5cm}{\includegraphics[width=4.5cm]{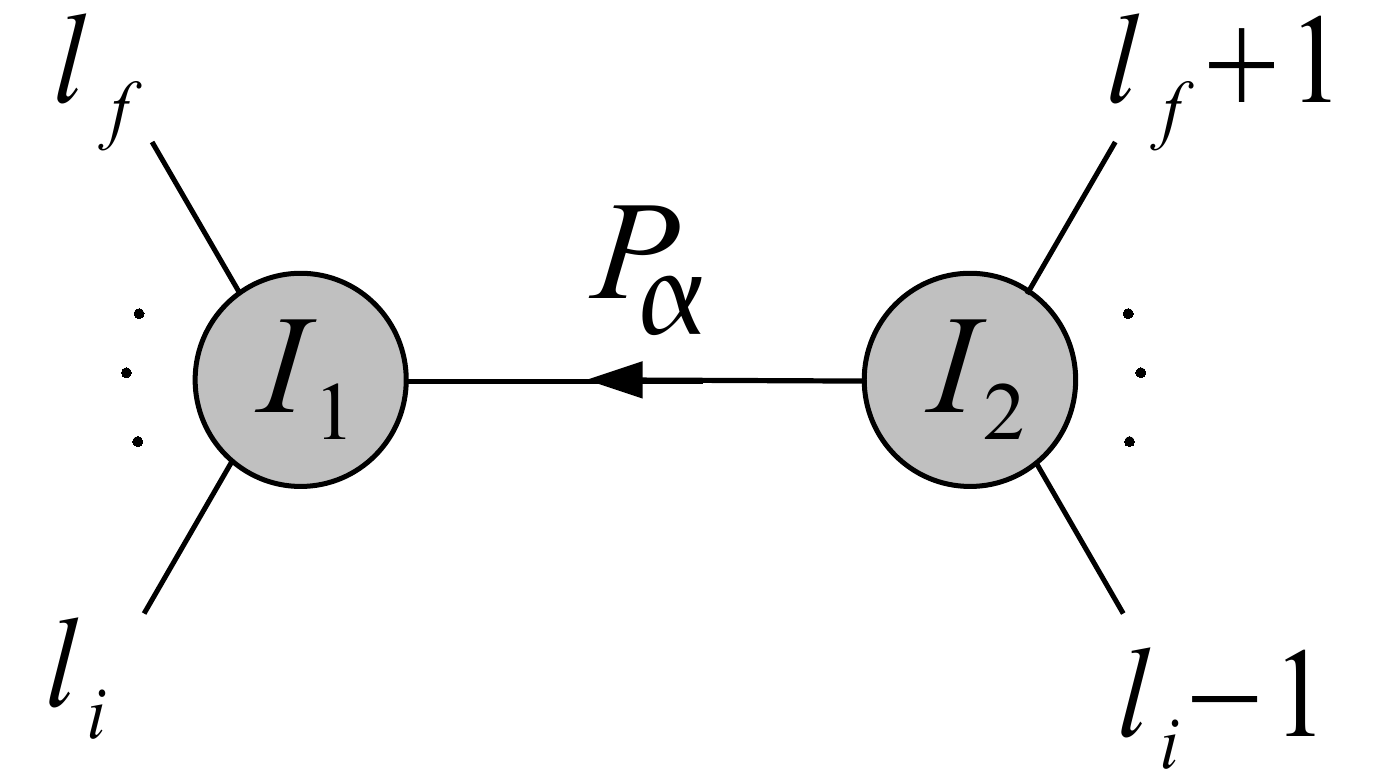}}\,.
\end{split}
\end{equation}
The sum in \reef{ANMHV} includes all products of MHV subamplitudes\footnote{We will often refer to subamplitudes as vertices.} in
which $I_1$ and $I_2$ contain $n_1 $ and $n_2 $ lines respectively, with
$n_1 + n_2 = n+2$ and $n_1,n_2 \ge 3$.
Further one must sum over the $n$ cyclically ordered assignments of external states $1, 2, \dots n$ to the lines $l_i,\dots l_f, l_f+1,\ldots,l_i-1.$   There are a total of  $n(n-3)/2$ distinct diagrams.

As the diagram in \reef{ANMHV} illustrates, the subamplitude $I_1$ contains  $n_1-1$ external states, which we denote as the set $\alpha=\{l_i,\ldots,l_f\}$, plus the state corresponding to the internal line $-P_\a$.
Similarly, the MHV subamplitude $I_2$ contains the remaining $n_2-1$ external states $\bar \alpha=\{l_f\!+\!1,\ldots,l_i\!-\!1\}$ and the internal state of momentum $P_\alpha$.   The direction chosen for $P_\a$ is indicated by an arrow, so that the internal momentum is given by
\begin{equation}\lab{chmom}
    P_\alpha=\sum_{i\in\alpha}p_i=p_{l_i}+\ldots+p_{l_f}\,.
\end{equation}
The sign of $P_\a$ is fixed by our convention that all external lines are outgoing.

The subamplitudes  $\MHV (I_1)$  and $\MHV (I_2)$ are
easily computed from \reef{mhvgen} and take the form \reef{genmhv}. The momentum assignments are
\begin{equation}   \lab{vert}
    \MHV (I_1)=\MHV (l_i,\ldots,l_f,-P_{\alpha})\,,\qquad
    \MHV (I_2)=\MHV (P_{\alpha},l_f+1,\ldots,l_i-1)\,.
\end{equation}
The internal momentum $P_\a$ is not a null vector. Instead the CSW prescription \cite{csw} instructs us to use the spinors
$|\pm P_\a\>$ defined by
\begin{equation}\label{CSWp}
    |\pm P_{\alpha}\>\,\equiv\, \pm  P_{\alpha}|X]\,=\, \pm\sum_{i\in\alpha}|i\>[iX]\,
\end{equation}
for the internal states in \reef{vert}.
(This is discussed further in section \ref{secVEfromCIS}.)
The reference spinor $|X]$ is arbitrary. Individual diagrams depend on the choice of $|X]$, but their sum reproduces the correct on-shell amplitude which is  independent of $|X]$. This is an important consistency requirement on the MHV vertex expansion.  We illustrate the $|X]$ independence of amplitudes in section~\ref{sec2Xindep} with  examples.

It is useful to outline the way the requirement of $SU(4)$ symmetry determines the diagrams which are non-vanishing in the expansion  \reef{ANMHV} for a
specific choice of NMHV amplitude. As discussed in section~\ref{secnotation}, the external
states $1,2,\ldots,n$ carry a total of 12 upper $SU(4)$ indices, and each index value
$1,2,3,4$ must appear 3 times.  In any given diagram these indices are shared
by the states in the sets $\a$ and $\bar{\a}$.  The subamplitude $I_1$
is nonvanishing only if
each index value appears either once or twice on the external states in $\a$. In particular, the states in $\a$ cannot contain more than $8$ or less than $4$ indices.
For non-vanishing amplitudes
there is then a unique
assignment of indices to the state $-P_\a$ such that subamplitude $I_1$
is $SU(4)$ invariant with a total of 8 indices and each index value appearing
twice.  In this case the subamplitude $I_2$ is also  $SU(4)$ invariant.  Note
that if the state $-P_\a$ is emitted from subamplitude
$I_1$
as a particle carrying
a subset of the 4 possible index values, then the state $P_\a$ is the
 anti-particle emitted from subamplitude $I_2$
and carries the complementary subset of indices.  Thus there are
a total of 4 distinct $SU(4)$ indices associated with the internal line.  The
Wick contraction algorithm discussed in section~\ref{secnotation} shows that spin factors of the MHV  subamplitudes $I_1$ and $I_2$ are also quickly determined from the $SU(4)$ indices on the lines they contain.   Thus  a great deal of quick information
can be obtained from the configuration of $SU(4)$ indices.

\example
Consider the 6-point amplitude $\<A^1(1)A^{23}(2)A(3)A^{234}(4)A^{14}(5)A^{1234}(6)\>$.   The diagram in which $I_1$ contains the states
$\a=\{1,2,3\}$
vanishes because the index value 4 is not present  among those states.   On the other hand, the diagram for the channel $\a=\{4,5\}$ does not
vanish. Its internal line $-P_\a$ carries indices $123$.
This diagram can then be computed as
\be \lab{ugly}
\parbox[c]{4.5cm}{\includegraphics[width=4.5cm]{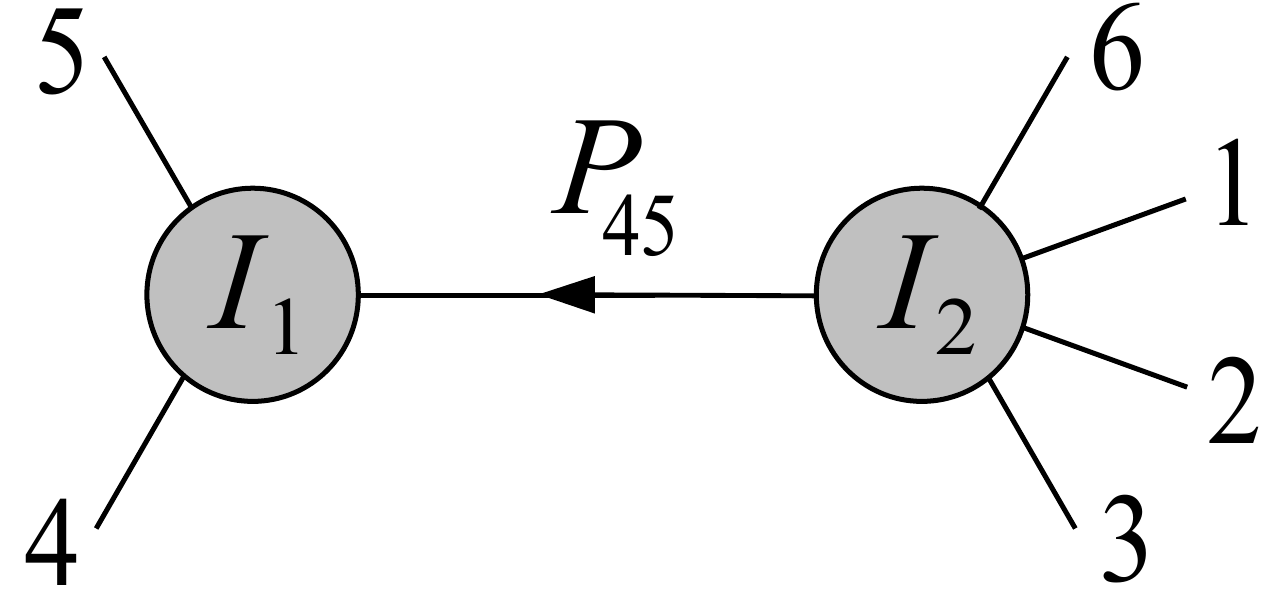}}
\,=~  \frac{1}{P^2_{45}}
\Biggl[\frac{\<45\>\<5,-P_{45}\>\<4,-P_{45}\>^2}{\<45\>\<5,-P_{45}\>\<-P_{45},4\>}\times\frac{\<16\>\<26\>^2\<6 P_{45}\>}{\<61\>\<12\>\<23\>\<3P_{45}\>\<P_{45}6\>}\Biggr]\,.
\ee
Further consideration of the $SU(4)$ indices quickly tells us that the only non-vanishing diagrams are those for the four channels $\a=\{4,5\},~\{5,6\},~\{6,1\},~\{3,4,5\}$.

We now discuss the extension of this construction to $k > 1$.
 MHV vertex diagrams for N$^k$MHV amplitudes contain $k$ internal lines.
We choose an arrow to indicate the direction of each internal momentum
$P_\a$.  By convention we define the channel $\a$ as the
set of all external lines on the side of the diagram that the arrow points towards.
Each channel could also have been defined by the complementary set
of external lines, which we will denote $\bar{\a}$. The channel momentum is given by
\be \lab{chmom}
P_{\a} \,=\, \sum_{i \in \a} p_i\,=\, -\sum_{j\in \bar{\a}} p_j
\,=\, - P_{\bar{\a}}\,.
\ee
Every MHV subamplitude contains one or more internal lines.  In
every case  one uses  the CSW prescription \reef{CSWp} for the angle spinor $|\pm P_\a\>.$  As in \reef{ugly},  the spinors $|\pm P_\a\>$ for each internal state always occur
four times in the denominator and four times in the numerator's  product of spin factors of the two adjacent subamplitudes.

For N$^2$MHV amplitudes ($k=2$) there is only one type of MHV vertex diagram, which contains  the product of
three MHV subamplitudes $I_1$, $I_2$, and $I_3$ and two internal propagators. As indicated in \reef{AN2MHv} below,
the two internal lines $\alpha_1, \alpha_2$ are described as the channels $\alpha_1=\{l_i,\ldots,l_f\}$ and $\alpha_2=\{r_i,\ldots,r_f\}$.
The expansion includes all diagrams in which the subamplitudes $I_B$
($B=1,2,3$)
have
$n_B$ lines each, with $n_1+n_2+n_3 = n +4$ and $n_B \ge 3$.  Thus
\begin{equation}\label{AN2MHv}
\begin{split}
    \A^{\rm N^2MHV}_n(1,\ldots,n)
    &=\hskip-.0cm\sum_{\stackrel{\text{MHV vertex diagrams }}{\{\alpha_1,\alpha_2\}}}\hskip-.1cm
    \frac{\MHV \!(I_1)\MHV \!(I_2)  \MHV \!(I_3)}{P_{\alpha_1}^2 P_{\alpha_2}^2}\\[.3ex]
    &=\sum_{\stackrel{\text{MHV vertex diagrams }}{\{\alpha_1,\alpha_2\}}}
   \parbox[c]{7cm}{\includegraphics[width=7cm]{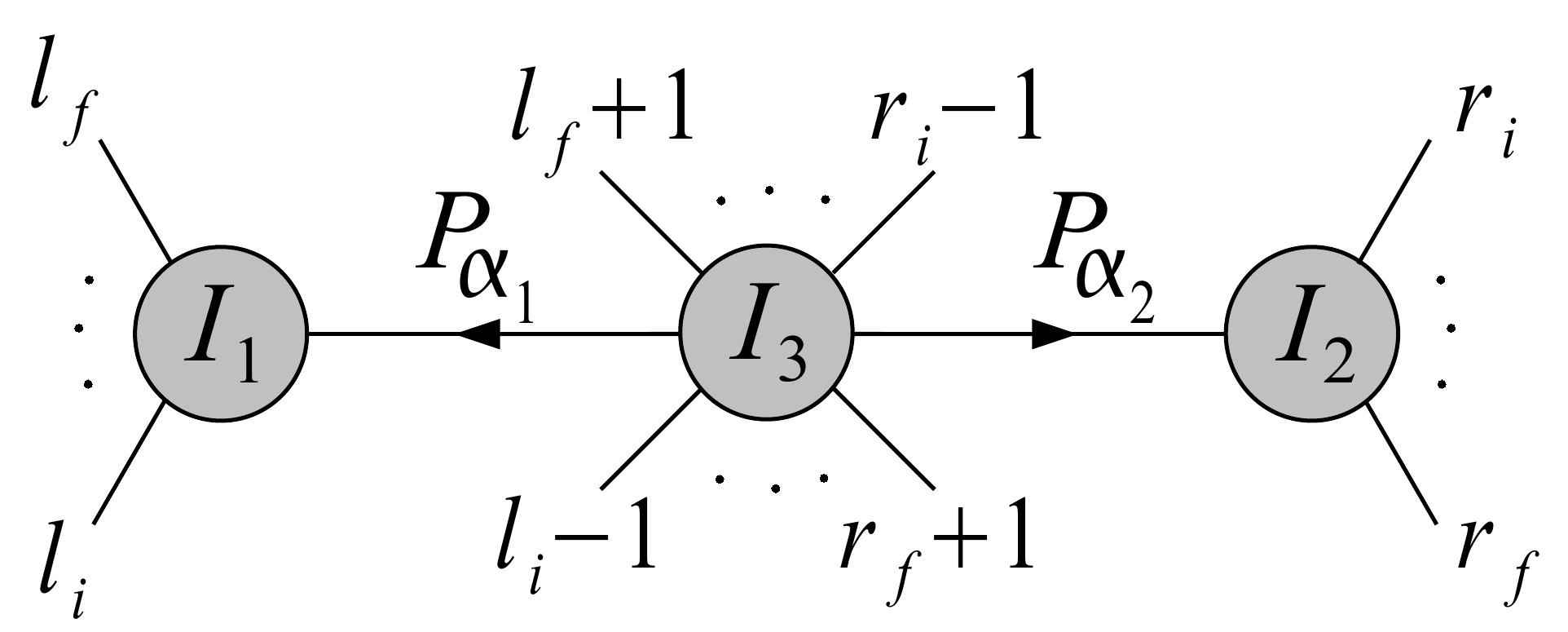}}~~~.
\end{split}
\end{equation}
The MHV subamplitudes corresponding to $I_1$, $I_2$, and $I_3$ can be read  directly from the diagram in~(\ref{AN2MHv}).   They are
\begin{equation}
\begin{split}
    \MHV (I_1)&=\MHV (l_i,..,l_f,-P_{\alpha_1})\,,\quad
    \MHV (I_2)=\MHV (-P_{\alpha_2},r_i,..,r_f)\,,\\[2mm]
    \MHV (I_3)&=\MHV \!(P_{\alpha_1}, l_f\!+\!1,.., r_i\!-\!1,P_{\alpha_2},r_f\!+\!1,..,l_i\!-\!1)\,.
\end{split}
\end{equation}

\example
Consider the MHV vertex diagrams in the expansion of  N$^2$MHV $8$-point amplitudes. There are $16$ ``skeleton'' diagrams, which are illustrated in figure~\ref{figeightpoint}.
\begin{figure}
\begin{center}
 \includegraphics[width=15cm]{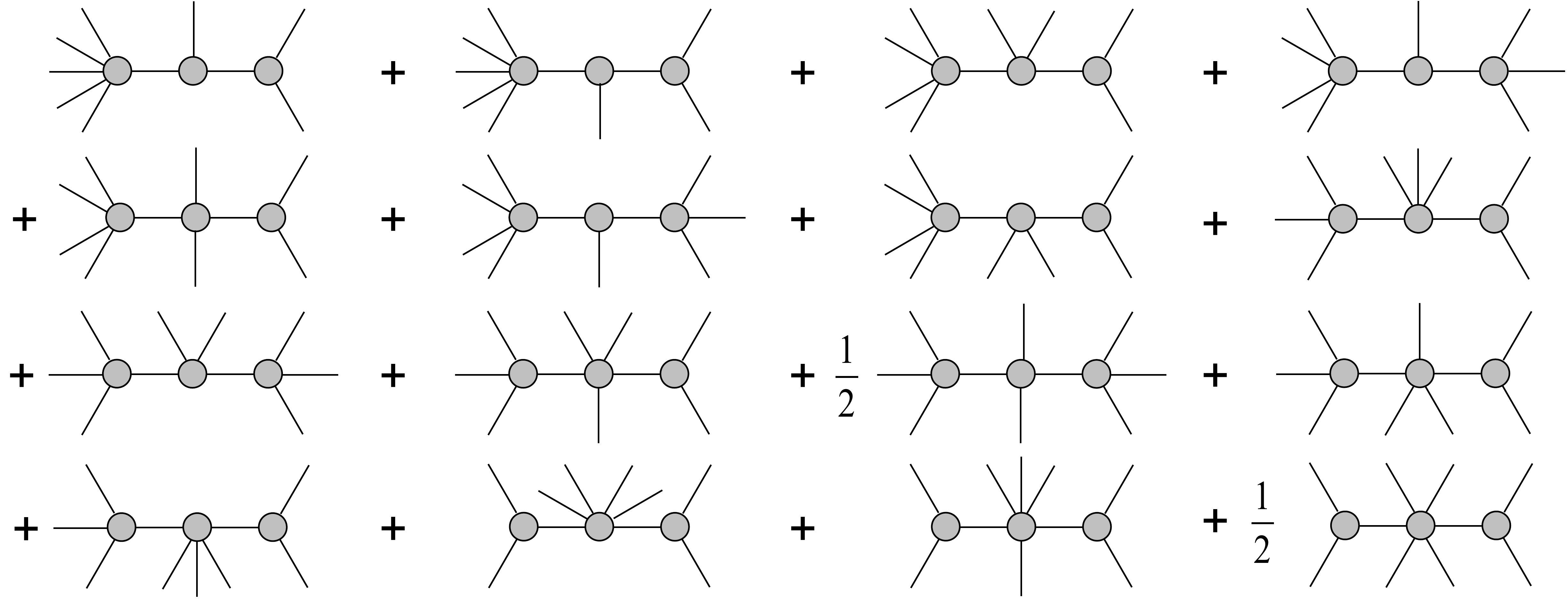}\\[3mm]
\end{center}
\vskip -.7mm
\caption{The 16 skeletons contributing to the MHV vertex expansion of $8$-point amplitudes at the N$^2$MHV level. Two diagrams carry a symmetry factor of $\frac{1}{2}$ because the sum over cyclic assignments of external particles to the legs of these skeletons overcounts their contribution. In fact, these
 two
diagrams are invariant under the permutation $i\to i+4$ of external legs.}
\lab{figeightpoint}
\end{figure}
To obtain the full set of diagrams from the skeletons, one needs to sum over the
distinct cyclic assignments of external states to the legs of the skeleton.
As for Feyman diagrams, some MHV vertex diagrams
are symmetric and one must include a symmetry factor to  compensate for the overcounting in the sum over cyclic permutations. In the 8-point example of figure~\ref{figeightpoint}, two skeletons
carry such a symmetry factor of $\frac{1}{2}$.

The number of skeletons needed for N$^2$MHV amplitudes grows as $O(n^3)$ in the number of external legs $n$. A short combinatoric analysis of the number of N$^2$MHV skeletons yields
\begin{equation}\label{skeletons}
  \frac{1}{12}n(n-2)(n-4)~~\text{ for even $n$}\,,\qquad
  \frac{1}{12}(n+1)(n-3)(n-4)~~\text{ for odd $n$}\,.
\end{equation}
The number of total diagrams can easily be obtained from the number of skeletons. Taking symmetry factors (which only occur for even $n$) into account,  we obtain
$n(n+1)(n-3)(n-4)/12$ as the number of N$^2$MHV diagrams for both even and odd $n$.

For $n=8$, eq.~(\ref{skeletons}) indeed gives
16 skeletons as displayed in figure~\ref{figeightpoint}. We have used these skeletons to implement the MHV vertex expansion for eight-point N$^2$MHV amplitudes in a \emph{Mathematica} program. The sum over cyclic permutations and the calculation of
spin factors is automated in the program.
We calculated several amplitudes numerically and confirmed that the results are independent of the choice of reference spinor $|X]$.

For N$^3$MHV amplitudes ($k=3$), there are two types of MHV vertex diagrams.
The first type is a linear chain of four MHV subamplitudes, which we call  a ``chain" diagram (see the figure in \reef{AN3MHVchain} below).
It contains  three internal lines $\alpha_1$, $\alpha_2$, and $\alpha_3$ for the channels $\alpha_1=\{l_i,\ldots,l_f\}$, $\alpha_2=\{m_i,\ldots,m_f\}$, and $\alpha_3=\{r_i,\ldots,r_f\}$.
Note that all lines in $\alpha_1$ are also contained in $\alpha_2$, \ie that $\alpha_1\subset \alpha_2$, and that $\alpha_3$ does not share any lines with $\alpha_1$, $\alpha_2$.
The ``chain'' contribution to the MHV vertex expansion of an amplitude $\A^{\rm N^3MHV}_{n}(1,\ldots,n)$  is the sum over all such diagrams, viz.
\begin{equation}\label{AN3MHVchain}
\begin{split}
    \A^{\rm N^3MHV}_{n,\,{\rm chain}}(1,\ldots,n)
    &=\hskip-.3cm\sum_{\stackrel{\text{``chain'' diagrams }}{\{\alpha_1,\alpha_2,\alpha_3\}}}\hskip-.1cm
    \frac{\MHV (I_1)\MHV (I_2)\MHV (I_3)\MHV (I_4)}{P_{\alpha_1}^2 P_{\alpha_2}^2P_{\alpha_3}^2}\\
    &=\hskip-.3cm\sum_{\stackrel{\text{``chain'' diagrams }}{\{\alpha_1,\alpha_2,\alpha_3\}}}
    \parbox[c]{9.5cm}{\includegraphics[width=9.5cm]{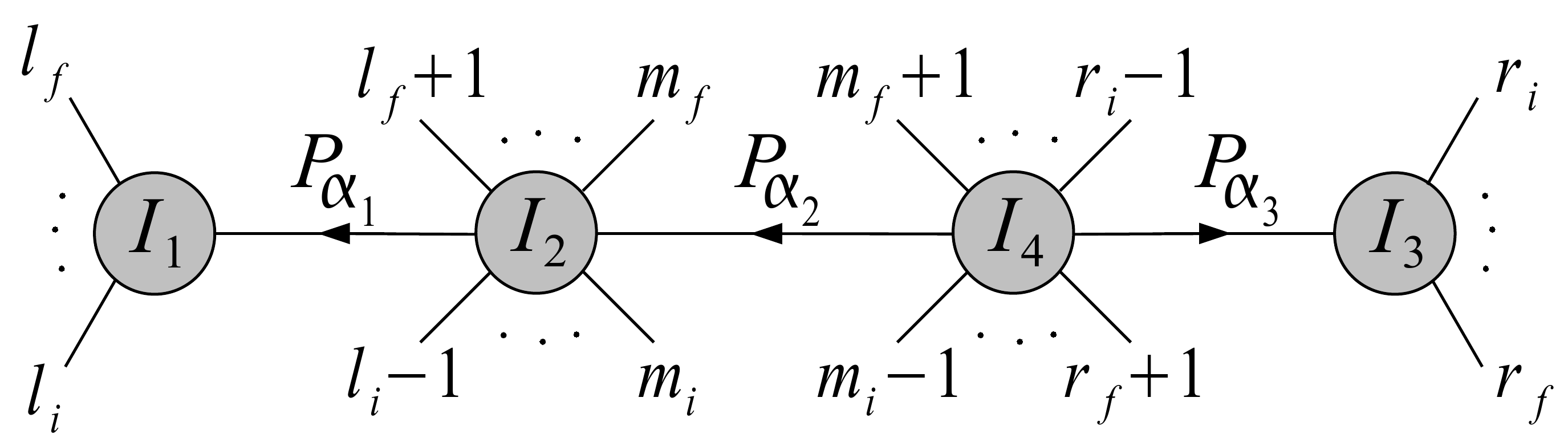}}~.
\end{split}
\end{equation}

The second type of MHV vertex diagram which contributes to the MHV vertex expansion of N$^3$MHV amplitudes contains one central MHV subamplitude which is connected through the internal lines $\alpha_1$, $\alpha_2$, and $\alpha_3$ to the three remaining MHV subamplitudes.
We  refer to this as a ``star'' diagram. The channels by $\alpha_1=\{l_i,\ldots,l_f\}$, $\alpha_2=\{t_i,\ldots,t_f\}$, and $\alpha_3=\{r_i,\ldots,r_f\}$ are defined so
that they do not share any external lines.
The ``star'' contribution to the MHV vertex expansion of an amplitude $\A^{\rm N^3MHV}_{n}(1,\ldots,n)$ is then given by
\begin{equation}\label{AN3MHVstar}
    \A^{\rm N^3MHV}_{n,\,{\rm star}}(1,\ldots,n)
    ~=~\sum_{\stackrel{\text{``star'' diagrams }}{\{\alpha_1,\alpha_2,\alpha_3\}}}~~
    \parbox[c]{6cm}{\includegraphics[width=6cm]{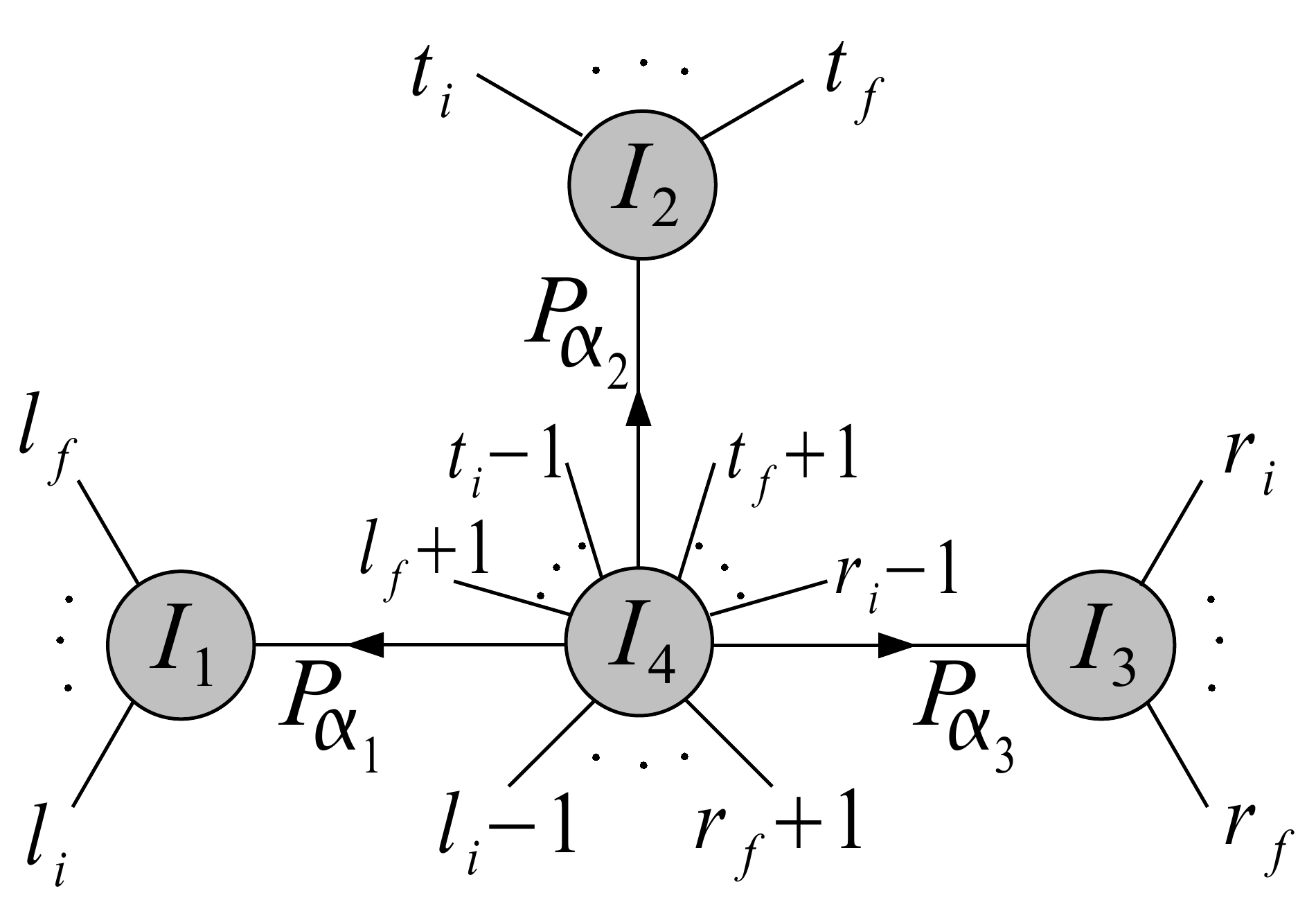}}~~~~.
\end{equation}
Adding the ``chain'' and the ``star'' contributions, we  write the complete MHV vertex expansion of any N$^3$MHV amplitude:
\begin{equation}
\begin{split}
     \A^{\rm N^3MHV}_{n}(1,\ldots,n)
     ~=~&\A^{\rm N^3MHV}_{n,\,{\rm chain}}(1,\ldots,n)+\A^{\rm N^3MHV}_{n,\,{\rm star}}(1,\ldots,n)\,.
\end{split}
\end{equation}

Consider now a general N$^k$MHV $n$-point amplitude $\A^\text{N$^k$MHV}_n(1,\ldots,n)$. There are several topologically distinct types of diagrams. Each diagram
 is characterized by a set of channels $\{\alpha_1,\ldots,\alpha_k\}$ which correspond to its internal lines.
There are $k+1$ MHV subamplitudes $I_1,\ldots,I_{k+1}$, and each external state appears in exactly one  subamplitude. Each internal line $\alpha_A$ appears once as a state of momentum $P_{\alpha_A}$ and once as a state of momentum $-P_{\alpha_A}$. The MHV vertex expansion of the amplitude $\A^\text{N$^k$MHV}_n(1,\ldots,n)$ is then given by
\begin{equation}\label{AnNkMHV0}
    \A^\text{N$^k$MHV}_n(1,\ldots,n) =\sum_{\stackrel{\text{MHV vertex diagrams }}{\{\alpha_1,\ldots,\alpha_k\}}}
    \frac{\MHV (I_1)\cdots \MHV (I_{k+1})}{P_{\alpha_1}^2\cdots P_{\alpha_k}^2} \, .
\end{equation}
The sum includes all sets of channels $\{\alpha_1,\ldots,\alpha_k\}$ for which
the subamplitudes $I_B$ are each MHV (and thus each carry 8 $SU(4)$ indices
with each index value occurring twice).
The CSW prescription~(\ref{CSWp}) is understood for all occurrences of angle spinors $|P_{\alpha_A}\>$ and $|-P_{\alpha_A}\>$ in the expressions for the subamplitudes $\MHV (I_1),\ldots,\MHV (I_{k+1})$.

 \subsection{$|X]$-independence and examples}
\label{sec2Xindep}
Each non-vanishing diagram in the MHV vertex expansion of an amplitude $\A^\text{N$^k$MHV}_n(1,\ldots,n)$ typically depends on the reference spinor $|X]$. The full amplitude must be  independent of the choice of $|X]$.
Therefore the dependence on $|X]$ must cancel after summing all contributing diagrams. We now illustrate this property with the computation of the seven-point N$^3$MHV amplitude
\begin{equation}\label{2diagramexample}
    \A^{\rm N^3MHV}_7(1,2,3,4,5,6,7) =  \bigl\<\,A^{12}(1)\,A^{1234}(2)\,A^{23}(3)\,A^{234}(4)\,A^{134}(5)\,A^{134}(6)\,A^{124}(7)\,\bigr\> \, .
\end{equation}
We choose a seven-point amplitude because an anti-MHV calculation permits an independent check of our result. The chosen amplitude~(\ref{2diagramexample}) has the simplifying feature that only two MHV vertex diagrams contribute to its expansion. Both diagrams are of the ``star'' type \reef{AN3MHVstar}. The first diagram is characterized by the internal lines
\begin{equation}
    \alpha_1=\{1,2\}\,,\qquad \alpha_2=\{4,5\}\,,\qquad \alpha_3=\{6,7\}\qquad~~\text{(\,figure \ref{figMHVVD3}a\,)}\,.
\end{equation}
\begin{figure}
\begin{center}
 \includegraphics[width=15cm]{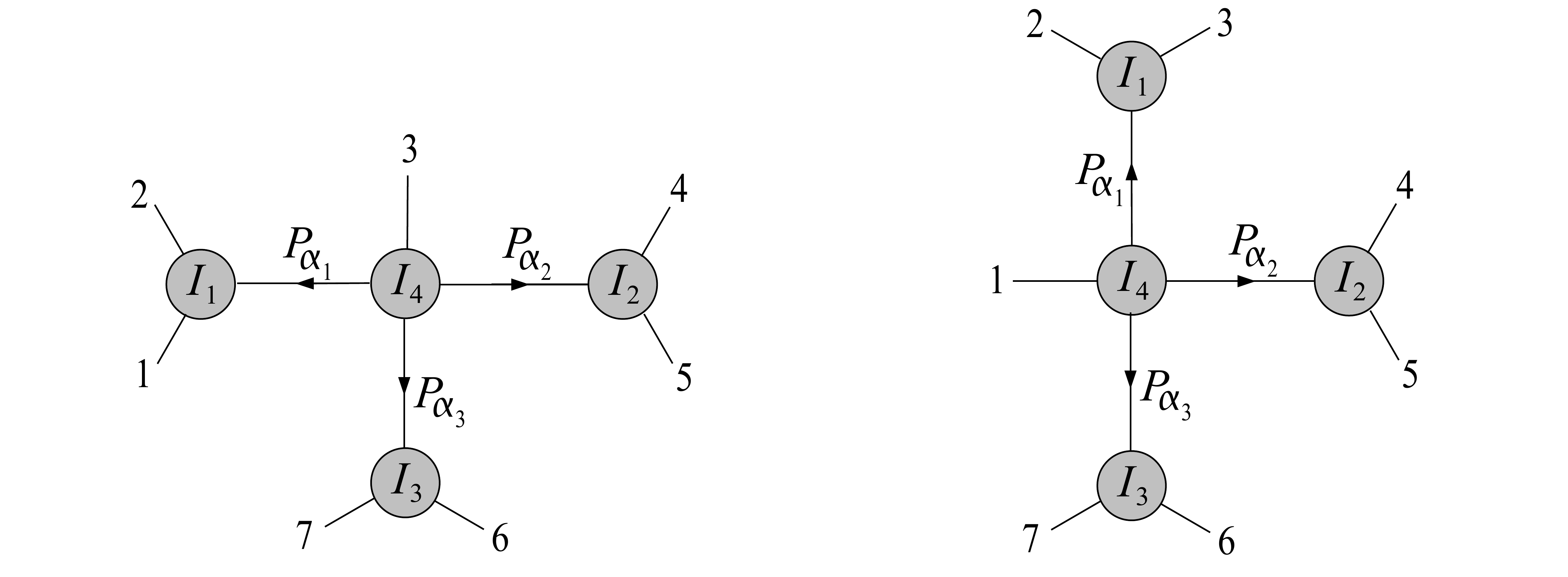}\\[3mm]
\end{center}
\vskip -.7mm
\centerline{(a)\hskip6.5cm(b)\hskip.25cm}
\caption{The two vertex diagrams that contribute to the MHV vertex expansion of the seven-point  N$^{3}$MHV amplitude $\<A^{12}(1)A^{1234}(2)A^{23}(3)A^{234}(4)A^{134}(5)A^{134}(6)A^{124}(7)\>$.}
\lab{figMHVVD3}
\end{figure}
Its contribution is given by
\begin{equation}\label{diag1}
\begin{split}
    \frac{\MHV (I_1)\!\times\! \MHV (I_2)\!\times\! \MHV (I_3)\!\times\! \MHV (I_4)}{P_{12}^2P_{45}^2P_{67}^2}
    &=\frac{1}{P_{12}^2P_{45}^2P_{67}^2}\Biggl[\frac{\<12\>\<2P_{12}\>}{\<P_{12}1\>}  \times\<45\>    \times\<67\>  \times 1\Biggr]\\
    &=-\frac{\<2|P_{12}|X]}{[12][45][67]\<1|P_{12}|X]}
    = \frac{[1X]}{[12][45][67][2X]}\,.
\end{split}
\end{equation}
We used the CSW prescription to make the dependence on the reference spinor $|X]$ explicit, and simplified the spinor products using $\<2|P_{12}|X]=\<21\>[1X]$.
The dependence on $|P_{45}\>$ and $|P_{67}\>$ canceled completely between spin factors in the numerator and the cyclic product of angle brackets in the denominator of the MHV subamplitudes $I_2$ and $I_3$. This cancelation is related to the particular simplicity of the MHV vertex expansion of this amplitude.
The second diagram is characterized by the internal lines
\begin{equation}
    \alpha_1=\{2,3\}\,,\qquad \alpha_2=\{4,5\}\,,\qquad \alpha_3=\{6,7\}\qquad~~\text{(\,figure \ref{figMHVVD3}b\,)}\,.
\end{equation}
For its contribution, we readily obtain
\begin{equation}\label{diag2}
\begin{split}
    \frac{\MHV (I_1)\!\times\! \MHV (I_2)\!\times\! \MHV (I_3)\!\times\! \MHV (I_4)}{P_{23}^2P_{45}^2P_{67}^2}
    &=\frac{1}{P_{23}^2P_{45}^2P_{67}^2}\Biggl[\frac{\<23\>\<P_{23}2\>}{\<3P_{23}\>}  \times\<45\>    \times\<67\>  \times 1\Biggr]\\
    &=-\frac{\<2|P_{23}|X]}{[23][45][67]\<3|P_{23}|X]}
    = \frac{[3X]}{[23][45][67][2X]}\,.
\end{split}
\end{equation}
The sum of the two contributions~(\ref{diag1}) and~(\ref{diag2})
  is the full amplitude. It simplifies nicely by the Schouten identity:
\begin{equation}
\begin{split}
    \A^{\rm N^3MHV}_7(1,2,3,4,5,6,7)
    &=\frac{[1X]}{[12][45][67][2X]}+\frac{[3X]}{[23][45][67][2X]}
    =\frac{[23][1X]+[12][3X]}{[12][23][45][67][2X]}\\[1mm]
    &=\frac{[13]}{[12][23][45][67]}
    = - \frac{[13][17][34][56]}{[12][23][34][45][56][67][71]}\,.
\end{split}
\end{equation}
This result is indeed independent of the reference spinor $|X]$ and agrees with the expected anti-MHV expression displayed above.
\medskip

Although many diagrams contribute to the MHV vertex expansion of a typical amplitude, it is curious that there are amplitudes which require only  a single MHV vertex diagram. In fact, it is easy to construct such amplitudes explicitly at the NMHV, N$^2$MHV, and N$^3$MHV level for any number of external states. One example is the class of non-vanishing N$^3$MHV amplitudes
\begin{equation}\label{1diagramexample}
    \A^{\rm N^3MHV}_n =  \bigl\<\,A^{134}(1)\,\ldots\,A^{124}(s)\,A^{124}(s+1)\,\ldots\,A^{123}(t)\,A^{123}(t+1)\,\ldots\,A^{234}(n-1)\,A^{34}(n)  \,\bigr\>\,.
\end{equation}
Here, the dots $\ldots$ denote an arbitrary number of positive helicity gluon states. The unique MHV vertex diagram is of the ``star'' type with channels $\alpha_1=\{1,\ldots,s\}$, $\alpha_2=\{s+1,\ldots,t\}$, and $\alpha_3=\{t+1,\ldots,n-1\}$. The value of this diagram
must be independent of $|X]$, and
it is.

\setcounter{equation}{0}
\section{Generating functions for N$^k$MHV amplitudes}\label{secgenfct}

We now discuss the construction of  generating functions $\cf^\text{N$^k$MHV}_n$ which package all N$^k$MHV amplitudes
in a single expression.  These generalize the MHV generating function  \reef{mhvgen}.  There are $4n$ Grassmann variables
$\eta_{ia}$ associated with the external lines.  We know that every amplitude
 $\A^\text{N$^k$MHV}_n(1,\ldots,n)$ implicitly carries a set of $4k+8$ $SU(4)$ indices which specify its external states.  For each set of states there is
 a   unique  Grassmann differential operator $D^{(4k+8)}$ of order $4k+8$
 obtained  through the correspondence \reef{diffop}.
Any desired amplitude is obtained by  applying $D^{(4k+8)}$ to the generating function $\cf^\text{N$^k$MHV}_n$.

The strategy to construct the generating function from the MHV vertex expansion \reef{AnNkMHV0}  is quite straightforward.  In addition to the
$\h_{ai}$ associated with external lines we now include a Grassmann variable
$\eta_{\alpha_{\!A} a}$ for each internal line $\alpha_A$
and define the 4th order differential operator
\begin{equation}
    D^{(4)}_{\alpha_A}\equiv \prod_{a=1}^4\frac{\pa}{\pa \eta_{\alpha_{\!A}a}}\,.
\end{equation}
The expansion  \reef{AnNkMHV0}
of any amplitude $\A^\text{N$^k$MHV}_n(1,\ldots,n)$ can then be written as
\begin{equation}\label{AnNkMHV}
    \A^\text{N$^k$MHV}_n(1,\ldots,n) =D^{(4k+8)}\sum_{\stackrel{\text{MHV vertex diagrams }}{\{\alpha_1,\ldots,\alpha_k\}}}
    \Biggl[\,\prod_{A=1}^{k}D^{(4)}_{\alpha_A}\Biggr]
    \frac{\cf^{\rm MHV}(I_1)\cdots \cf^{\rm MHV}(I_{k+1})}{P_{\alpha_1}^2\cdots P_{\alpha_k}^2}\,.
\end{equation}
Each MHV subamplitude $\MHV (I_B)$ in  \reef{AnNkMHV0}  has been replaced by its generating function
\begin{equation}\label{cfMHV}
    \cf^{\rm MHV}(I_B)=\frac{\delta^{(8)}(I_B)}{\cyc (I_{B})} \,.
\end{equation}
The factors $\delta^{(8)}(I_B)$ and $ \cyc (I_{B})$ are defined as in \reef{dident} and \reef{cyc}, but they now include
the spinors  $|\pm P_{\a_A}\>$ and $\h_{\alpha_{\!A} a}$ for the internal states.
The justification for \reef{AnNkMHV} is keyed to the action of  the operator
$D^{(4k+8)}$ which selects the external states of the amplitude.  Their $SU(4)$ indices then uniquely determine the indices of the internal states
$\pm P_{\a_A}$ in each subamplitude.  The operator   $D^{(4)}_{\alpha_A}$  then splits uniquely into two
factors which correspond  to the required split of the $SU(4)$ indices on each
end of the internal line.

 The differential operator $D^{(4k+8)}$ now acts in \reef{AnNkMHV} on an expression that makes no reference to the specific amplitude under consideration.
 We therefore  define the generating function
\begin{equation}  \lab{gen1}
    \cf^\text{N$^k$MHV}_n(\eta_{1a},\ldots,\eta_{na}) =\sum_{\stackrel{\text{ all MHV vertex diagrams }}{\{\alpha_1,\ldots,\alpha_k\}}}\Biggl[\,\prod_{A=1}^{k}D^{(4)}_{\alpha_A}\Biggr]
    \frac{\cf^{\rm MHV}(I_1)\cdots \cf^{\rm MHV}(I_{k+1})}{P_{\alpha_1}^2\cdots P_{\alpha_k}^2}\,,
\end{equation}
from which  \emph{any} N$^k$MHV amplitude can be computed
by acting with the associated differential operator of order   $D^{(4k+8)}$.  Substituting the MHV generating functions from~(\ref{cfMHV}), we obtain
\begin{equation}\label{cfNkMHVraw}
    \cf^\text{N$^k$MHV}_n(\eta_{1a},\ldots,\eta_{na})=\sum_{\stackrel{\text{all MHV vertex diagrams }}{\{\alpha_1,\ldots,\alpha_k\}}}\Biggl[\,\prod_{A=1}^{k}D^{(4)}_{\alpha_A}\Biggr]
    \frac{\delta^{(8)}(I_1)\cdots \delta^{(8)}(I_{k+1})}{\cyc (I_1)\,\cdots\,\cyc (I_{k+1}) ~ P_{\alpha_1}^2\cdots P_{\alpha_k}^2}\,.
\end{equation}

The next step is to simplify~(\ref{cfNkMHVraw}) by evaluating the
derivatives $D^{(4)}_{\alpha_A}$.
It is convenient to use the identity
\begin{equation}\label{deltaID}
    \prod_{B=1}^{k+1}\delta^{(8)}\Big(I_B\Bigr)=\delta^{(8)}\biggl(\,\sum_{i\in{\rm ext}}|i\>\eta_{ia}\biggr)\prod_{A=1}^{k}\delta^{(8)}\biggl(\,\sum_{i\in\alpha_A}|i\>\eta_{ia}-|P_{\alpha_A}\>\eta_{\alpha_{\!A}a}\biggr)\,.
\end{equation}
The sum $\sum_{i\in{\rm ext}}$ is over all external legs, $i=1,\dots,n$.
The proof of the identity \reef{deltaID} is given in appendix~\ref{appdeltaID}. Let us here illustrate (\ref{deltaID}) with a specific example. Consider an MHV vertex diagram of the form displayed in figure~\ref{figMHVVD4}.
\begin{figure}
\begin{center}
 \includegraphics[width=6cm]{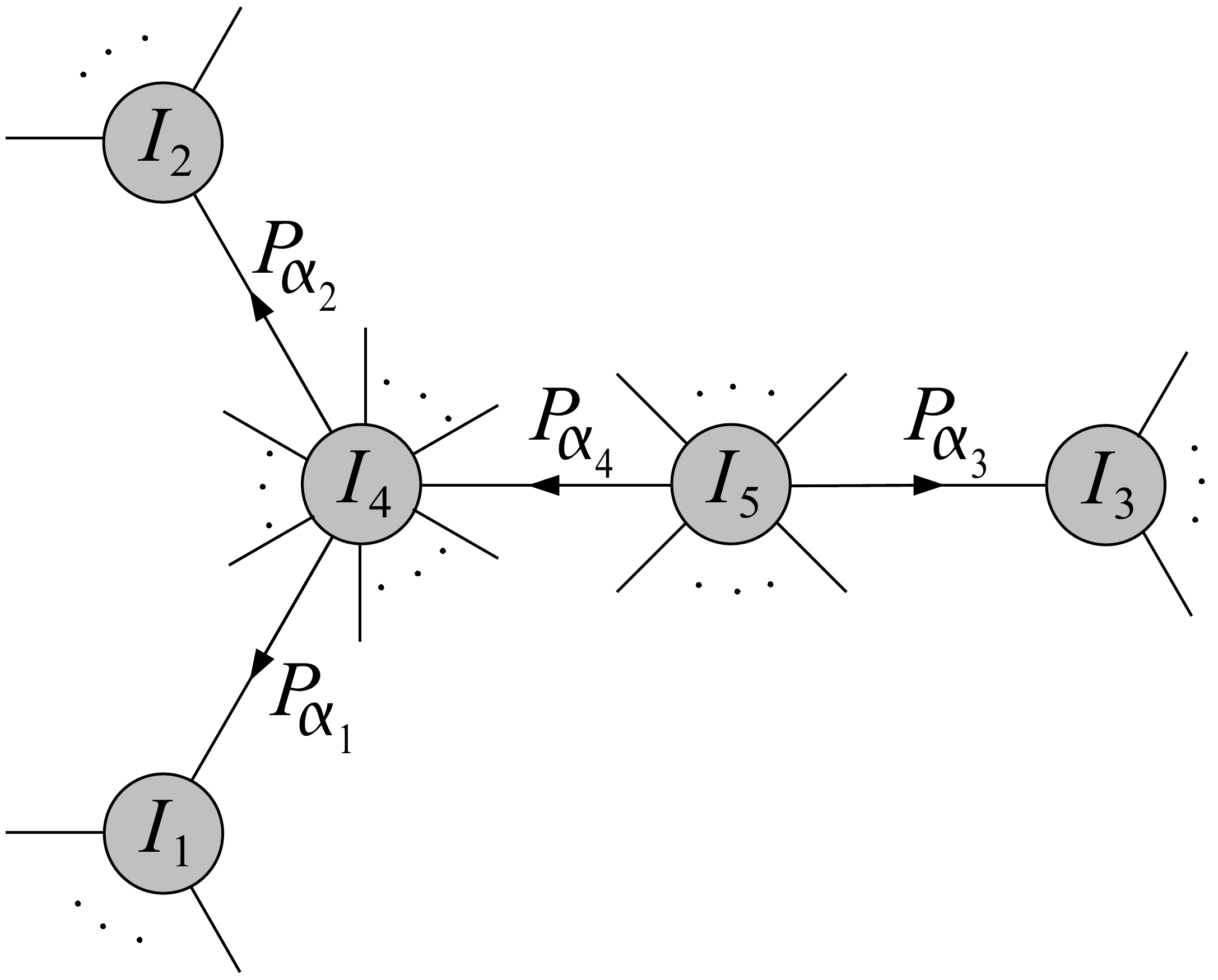}\\[3mm]
\end{center}
\vskip -.7mm
\caption{A typical diagram in the MHV vertex expansion of an  N$^{4}$MHV amplitude.}
\lab{figMHVVD4}
\end{figure}
Diagrams of this form can contribute to the MHV vertex expansion of N$^{4}$MHV amplitudes. From the diagram, we see that
\begin{equation}
\begin{split}
    \delta^{(8)}\Big(I_B\Bigr)&=\delta^{(8)}\biggl(\,\sum_{{\rm ext}\,i\,\in I_B}|i\>\eta_{ia}~-~|P_{\alpha_B}\>\eta_{\alpha_Ba}\biggr)
    \qquad\text{for }B=1,2,3~\,,\\
    \delta^{(8)}\Big(I_4\Bigr)&=\delta^{(8)}\biggl(\,\sum_{{\rm ext}\,i\,\in I_4}|i\>\eta_{ia}
    ~-~|P_{\alpha_4}\>\eta_{\alpha_4a}
    ~+~|P_{\alpha_1}\>\eta_{\alpha_1a}~+~|P_{\alpha_2}\>\eta_{\alpha_2a}\biggr)\,,\\
    \delta^{(8)}\Big(I_5\Bigr)&=\delta^{(8)}\biggl(\,\sum_{{\rm ext}\,i\,\in I_5}|i\>\eta_{ia}~+~|P_{\alpha_4}\>\eta_{\alpha_4a}
    ~+~|P_{\alpha_3}\>\eta_{\alpha_3a}\biggr)
\end{split}
\end{equation}
and
\begin{equation}\label{alphasexample}
    \alpha_1=\{\,{\rm ext}\,\,i\,\in I_1\,\}\,,\quad
    \alpha_2=\{\,{\rm ext}\,\,i\,\in I_2\,\}\,,\quad
    \alpha_3=\{\,{\rm ext}\,\,i\,\in I_3\,\}\,,\quad
    \alpha_4=\alpha_1\cup\alpha_2\cup\{\,{\rm ext}\,\,i\,\in I_4\,\}\,.
\end{equation}
It follows that
\begin{equation}\label{delta123}
\begin{split}
  &\delta^{(8)}\Big(I_1\Bigr)\,\delta^{(8)}\Big(I_2\Bigr)\,\delta^{(8)}\Big(I_3\Bigr)\,\delta^{(8)}\Big(I_4\Bigr)\,\delta^{(8)}\Big(I_5\Bigr)\\[.4ex]
  &=\delta^{(8)}\Big(I_1\Bigr)\,\delta^{(8)}\Big(I_2\Bigr)\,\delta^{(8)}\Big(I_3\Bigr)\,\delta^{(8)}\Big(I_1+I_2+I_4\Bigr)\,\delta^{(8)}\Big(I_1+I_2+I_3+I_4+I_5\Bigr)\\[.4ex]
  &=\delta^{(8)}\biggl(\,\sum_{i\in \alpha_1}|i\>\eta_{ia}~-~|P_{\alpha_1}\>\eta_{\alpha_1a}\biggr)
  \,\delta^{(8)}\biggl(\,\sum_{i\in \alpha_2}|i\>\eta_{ia}~-~|P_{\alpha_2}\>\eta_{\alpha_2a}\biggr)
  \,\delta^{(8)}\biggl(\,\sum_{i\in \alpha_3}|i\>\eta_{ia}~-~|P_{\alpha_3}\>\eta_{\alpha_3a}\biggr)\\[.4ex]
 &\quad\times \delta^{(8)}\biggl(\,\sum_{i\in \alpha_4}|i\>\eta_{ia}~-~|P_{\alpha_4}\>\eta_{\alpha_4a}\biggr)\,\,
 \delta^{(8)}\biggl(\,\sum_{i\in {\rm ext}}|i\>\eta_{ia}\biggr)\,,
\end{split}
\end{equation}
where we use the identity $\delta^{(8)}(X)\,\delta^{(8)}(Y)=\delta^{(8)}(X+Y)\,\delta^{(8)}(Y)$
repeatedly in the first step and then use~(\ref{alphasexample}) in the last step.
We have thus obtained the identity~(\ref{deltaID}) for the MHV vertex diagram displayed in figure~\ref{figMHVVD4}.

Carrying out the Grassmann differentiations $D^{(4)}_{\alpha_A}$ in the generating function~(\ref{cfNkMHVraw}) is now straightforward. For each
internal line we use
\begin{equation}
    D^{(4)}_{\alpha_A}\,\delta^{(8)}\biggl(\,\sum_{i\in\alpha_A}|i\>\eta_{ia}-|P_{\alpha_A}\>\eta_{\alpha_{\!A}a}\biggr)
    =\prod_{a=1}^4\sum_{i\in \alpha_A}\<i \, P_{\alpha_A}\>\h_{ia}\,.
\end{equation}
We immediately obtain
\begin{equation}\label{FnNkMHV}
\boxed{~~
\begin{split}
 &\cf^\text{N$^k$MHV}_n(\eta_{1a},\ldots,\eta_{na})
 =\sum_{\stackrel{\text{all MHV vertex diagrams }}{\{\alpha_1,\ldots,\alpha_k\}}}
 \frac{ \d^{(8)}\big(\sum_{i\in{\rm ext}}|i\>\h_{ia}\bigr)}{\cyc (I_1)\cdots\cyc (I_{k+1})} ~
 \prod_{A=1}^k\Biggl[\frac{1}{P_{\alpha_A}^2}\prod_{a=1}^4\sum_{i\in \alpha_A}\<i \,P_{\alpha_A}\>\h_{ia}\Biggr]\,.
\end{split}
}
\end{equation}
This concludes the derivation of the compact form for the N$^k$MHV generating function.

We conclude this section with some comments concerning the structure of the
generating function and how it is applied:
\begin{enumerate}
  \item
    We draw the reader's attention to a subtlety in the passage from
    \reef{AnNkMHV} to the generating function \reef{gen1}. The former is a rewrite of the MHV vertex expansion \reef{AnNkMHV0}
    for a specific amplitude
    and includes only the contribution of non-vanishing diagrams for which all subamplitudes $\MHV(I_A)$ are $SU(4)$ invariant.  However the generating function \reef{gen1} includes all possible MHV vertex diagrams that can be drawn for N$^k$MHV processes.
    For any specific amplitude  $\A^\text{N$^k$MHV}_n(1,\ldots,n)$,  the application of the associated  $D^{(4k+8)}$ produces
    a non-vanishing result
    only for the diagrams  which contribute to the expansion \reef{AnNkMHV0}.
  \item
  The final  form \reef{FnNkMHV} compactly summarizes
  all individual N$^k$MHV amplitudes.  One practical application of this
    form is the summation over the intermediate states in
    (generalized)
    unitarity cuts of loop amplitudes.  See \cite{efk1} for many examples at the NMHV level.  This application  involves products of
    several
    generating functions,  namely
    those for all subamplitudes resulting from the chosen cut.
  \item
  To evaluate a specific N$^k$MHV amplitude with $k=1,2$\,, it is usually easier to follow the method in our discussion
    of NMHV amplitudes in section~\ref{sec2diagrams}.  To obtain the explicit form of the diagrams which contribute to a particular amplitude, one simply writes down the  $SU(4)$
    indices for the external lines. The indices for the internal lines in any diagram are quickly and uniquely determined, and the product of spin factors is easily
    obtained. For large $k$, there are many diagrams and we expect a numerical implementation of the generating function~(\ref{FnNkMHV}) to be the more efficient method of computing amplitudes.

\end{enumerate}

\setcounter{equation}{0}
\section{The MHV vertex expansion from all-line shifts}\label{secVEfromCIS}
In this section we use \emph{all-line recursion relations} to derive the MHV vertex expansion for all tree amplitudes in $\cn=4$ SYM.
Our approach here generalizes the derivation of the MHV vertex expansion for gluon amplitudes in [\citen{risager},\citen{greatdane}], where a shift of only the $k+2$ negative helicity lines were used.
In section~\ref{secMHVVEcommon} we define all-line shifts, discuss their kinematics, and state the recursion relations they imply. The proof that these recursion relations are valid  will be presented in section~\ref{secvalidity}. In section~\ref{secMHVVE4line} we derive the MHV vertex expansion for N$^2$MHV amplitudes from the recursion relation of all-line shifts. We generalize our results to  N$^k$MHV amplitudes for all $k\geq1$ in section~\ref{secMHVVEall}.

\subsection{All-line shifts and their recursion relations}\label{secMHVVEcommon}
Consider any $n$-point N$^k$MHV amplitude $\A^\text{N$^k$MHV}_{n}(1,\ldots,n)$. We define an \emph{all-line shift} as the complex deformation of all square spinors
\begin{equation}\label{commindexshift}
  |i] ~\to~ |\hat i]=|i]+z\,c_i |X]\,,\hspace{1cm} i=1,\dots,n \, ,
\end{equation}
where $|X]$ is an arbitrary reference spinor. The complex parameters $c_i$ are constrained by momentum conservation, \ie
\begin{equation}\label{sumci}
    \sum_{i \in {\rm ext}} c_i|i\>=0\,.
\end{equation}
 Furthermore, we demand that an all-line shift satisfies momentum conservation only when \emph{all} external momenta are summed, as in~(\ref{sumci}). Namely, we demand
\begin{equation}\label{subsetsumci}
    \sum_{i\in \alpha}c_i|i\>\neq0
\end{equation}
for all proper subsets $\alpha$ of consecutive external lines.

\medskip
We will prove in section~\ref{secvalidity} that N$^k$MHV amplitudes in $\cn=4$ SYM
 with $k\geq1$
vanish as $z \to \infty$ under the all-line shift.
Therefore the all-line shift can be used to derive the valid recursion relation\begin{equation}\label{commonindexRR}
    \A^\text{N$^k$MHV}_{n}(1,\ldots,n)=\sum_{{\rm diagrams}\,\alpha} \A_{n_1}(\hat\alpha,-\hat P_\alpha)\,\frac{1}{P^2_\alpha}\,\A_{n_2}(\hat{\bar\alpha},\hat P_\alpha)\biggr|_{z=z_\alpha}\,,
\end{equation}
which is the starting point for our derivation of the MHV vertex expansion.
The subset of external lines on the subamplitudes $\A_{n_1}$ and $\A_{n_2}$ are denoted by $\alpha$  and $\bar{\a}$. The notation  $\hat{\a},~\hat{\bar{\a}},~\hat{P}_\a$ indicates that the momenta of the subamplitudes are shifted.
The conditions \reef{subsetsumci} ensure that all possible channels $\alpha$ can contribute, so the sum in \reef{commonindexRR} potentially includes all possible diagrams.\footnote{When the subamplitudes are not $SU(4)$ invariant, the corresponding diagram vanishes.}

There are $4k+8$ indices carried by  the external lines plus 4 more associated
with the internal states $\pm \hat{P}_\a$.  These $4k+12$ indices are shared
by the two subamplitudes in the combinations  $(8,\,4k+4), ~(12,\, 4k),\ldots, (4k+8,\,8)$. The combinations $(4,\,4k+8)$ and $(4k+8,\,4)$ are absent because the 3-point anti-MHV subamplitudes vanish due to the kinematics of the square spinor shift \cite{risager}, as we show below.
Thus only N$^{q}$MHV subamplitudes with
 $0\leq q < k$
can contribute in \reef{commonindexRR}.

 Shifted momenta are null vectors.  In particular $\hat{P}_\a $ is the null vector
 \be \lab{nullp}
 \hat{P}_\a\,=\, P _\a+ z_\a \sum_{i\in\alpha} c_i  |i\>[X|\,.
 \ee
 The value $z=z_\alpha$ of the complex shift parameter is determined by the requirement that  $(\hat P_\alpha)^2=0$. This fixes
\begin{equation}\lab{zalph}
    z_\alpha=\frac{P_\alpha^2}{\,\,\sum_{i\in\alpha} c_i\<i|P_\alpha|X]\,}\,.
\end{equation}
Since  $\hat{P}_\a $ is null we can write
\be \lab{mkwontlikeit}
\hat{P}_\a \,=\, |\hat{P}_\a\>[\hat{P}_\a|\,.
\ee
From $\hat{P}_\a |X] \,=\,     |\hat{P}_\a\>[\hat{P}_\a\,X]$, we obtain
\begin{equation}\label{deriveCSW}
    |\hat P_\alpha\>=\frac{ \hat P_\alpha|X]}{[\hat P_\alpha X]}=\frac{ P_\alpha|X]}{[\hat P_\alpha X]}\,.
\end{equation}
In sections~\ref{secMHVVE4line}
 and~\ref{secMHVVEall}
 we will show that  the factor $[\hat{P}_\a\, X]$
 always cancels in the product of the subamplitudes, so that \reef{deriveCSW} can be replaced by the CSW prescription \reef{CSWp}.

The fact that anti-MHV 3-point subamplitudes do not contribute can easily be seen, for example as follows.
Consider a 3-particle  subamplitude $\A_{3}(\hat{\a},-\hat{P}_\a)$
(or $\A_{3}(\hat{\bar{\a}},\hat{P}_\a)$) with consecutive external lines $j$ and $k$. The subamplitude depends on the square brackets  $[\hat j\,\hat{k}],\,[\hat{P}_\a\,\hat{j} ],\,[ \hat{P}_\a\,\hat{k}].$  Since $\hat{P}_\a = \hat j + \hat k$ is a null vector
 and $\<jk\>\neq 0$,
 the
 square bracket $[\hat j\,\hat{k}]$
 vanishes.
In bispinor form, we have
\be \lab{bsp}
\hat{P}_\a \,=\, |\hat{P}_\a\>[\hat{P}_\a|\,=\, | j \>[\hat j| + |k\>[ \hat{k}|\,.
\ee
Hence $|\hat{P}_\a\>[\hat{P}_\a \,\hat{j}] =0$ and
$ |\hat{P}_\a\>[\hat{P}_\a \,\hat k] =0$, so $[\hat{P}_\a\,\hat{j} ],\,[ \hat{P}_\a\,\hat{k}]$
also vanish.
Since any anti-MHV 3-point function
consists of a  product of four square brackets in the numerator and three in the
denominator, it must vanish.

It is convenient to re-express the recursion relation~(\ref{commonindexRR}) using generating functions.  We are working inductively in $k$, and  we can
assume that the MHV vertex expansion and its generating functions are established for $q < k$.   The discussion of section~\ref{secgenfct}  shows that the amplitudes
which appear in  \reef{commonindexRR}  can be obtained by applying
 the Grassmann differential operator  $D^{(4k+8)} D_\a^{(4)}$
 to the product of generating functions for these amplitudes.  The first factor consists of the derivatives associated with the external states and the second corresponds to
 the internal line of each term in the recursion relation.   We can thus rewrite
 \reef{commonindexRR} as
 \begin{equation}\label{rewritecommonindexRR}
    \A^\text{N$^k$MHV}_{n}(1,\ldots,n)=\frac{1}{2}D^{(4k+8)}\sum_{q=0}^{k-1} \sum_{\alpha}
    D^{(4)}_\alpha \frac{\cf^\text{N$^{q}$MHV}(\hat\alpha,-\hat P_\alpha;\eta_{\alpha a})\,
    \cf^\text{N$^{(k-q-1)}$MHV}(\hat{\bar\alpha},\hat P_\alpha;\eta_{\alpha a})}{P^2_\alpha}\biggr|_{z=z_\alpha}\,.
\end{equation}
The generating functions also depend on the $\h_{ia}$ variables for external lines. This is not explicitly indicated to simplify the notation.
The symmetry factor $\frac{1}{2}$ in~(\ref{rewritecommonindexRR}) is necessary because for each channel $\alpha$,
 we now also include the equivalent term with $\alpha\leftrightarrow\bar \alpha$ in the sum.

{}For $k=1$, \ie~NMHV amplitudes, the subamplitudes in \reef{rewritecommonindexRR} are both MHV, and the MHV vertex expansion can readily be obtained as in \cite{efk1,BEF}. For $k>1$, the recursion relation \reef{rewritecommonindexRR}
is not yet in the form of the MHV
vertex expansion, but we will obtain this expansion by further processing and induction on $k$.
We first illustrate this for N$^2$MHV amplitudes and then generalize the result to arbitrary $k$.

\subsection{N$^2$MHV amplitudes}\label{secMHVVE4line}
For $k=2$, (\ref{rewritecommonindexRR}) takes the form
\begin{equation}\label{AN2MHV}
    \A^{\rm N^2MHV}_{n}(1,\ldots,n)
    =D^{(16)}\sum_{\alpha}D^{(4)}_\alpha\frac{\cf^{\rm MHV}(\hat{\alpha},-\hat P_\alpha;\eta_{\alpha a})\cf^{\rm NMHV}(\hat{\bar\alpha},\hat P_\alpha;\eta_{\alpha a})\,}{P^2_\alpha}\biggr|_{z=z_\alpha}\,.
\end{equation}
The MHV and NMHV  generating functions which appear describe amplitudes with $n_1$ and $n_2 = n - n_1 +2$ lines, respectively.
Since MHV (NMHV) amplitudes must have at least 3 (5) lines,
 the sum includes
 all channels for which
 $(n_1,\,  n_2) =  (3,\,n-1), (4,\,n-2),\dots, (n-3,5).$

The MHV generating function in   \reef{AN2MHV}  is given by
\begin{equation}\label{cfMHValpha}
    \cf^{\rm MHV}(\hat{\alpha},-\hat P_\alpha;\eta_{\alpha a})=\frac{\delta^{(8)}\bigl(\sum_{i\in \alpha}|i\>\eta_{ia}-|\hat P_\alpha\>\eta_{\alpha a}\bigr)}{\cyc  (\hat I_1)}=  \cf^{\rm MHV}(\alpha,-\hat P_\alpha;\eta_{\alpha a})     \,.
\end{equation}
Since the external angle spinors $|i\>$ are not affected by the
square spinor shift,
we have replaced $\hat{\a}$ by $\a$ in the last equality of \reef{cfMHValpha}.
The ``hat" in the cyclic factor  $\cyc  (\hat I_1)$ indicates the effect of the shift on  $|\hat{P}_\a\>$, which is given by \reef{deriveCSW}.

We use  the generating function~(\ref{FnNkMHV}) with $k=1$ for $\cf^{\rm NMHV}$ in~(\ref{AN2MHV}). As mentioned above, its validity has already established \cite{efk1}.
We choose to express the NMHV generating function in terms of the \emph{same reference spinor $|X]$  as in the shift~(\ref{commindexshift})}.
The generating function $\cf^{\rm NMHV}$ contains a sum over channels denoted by $\b$, see Fig.~4b. We choose $\b$ to \emph{not} include line $\hat{P}_\a$ (which is an ``external state'' of the NMHV subamplitude).
Thus we write
\bea
  \nonumber
    \cf^{\rm NMHV}(\hat{\bar\alpha},\hat P_\alpha;\eta_{\alpha a})
    &=&\sum_{\beta} \frac{\delta^{(8)}\bigl(\sum_{i\in\bar\alpha}|i\>\eta_{ia}+|\hat P_\alpha\>\eta_{\alpha a}\bigr)}{\cyc (\hat I_2)\cyc (\hat I_3)\,\hat P^2_\beta(z_\alpha)}\prod_{a=1}^4\sum_{i\in\beta}\<i\,\hat P_\beta(z_\alpha)\>\eta_{ia}\\
    &\equiv&\sum_{\beta} \cf_\beta^{\rm NMHV}(\hat{\bar\alpha},\hat P_\alpha;\eta_{\alpha a})\,.
    \lab{cfNMHV00}
\eea
We use the notation $\hat{P}_\b(z_\a)$ because this internal momentum
 contains shifted external lines of \reef{commindexshift}. Note that
 this momentum is off-shell,
\be \lab{pbsq}
 \hat P^2_\beta(z_\alpha)\,= P^2_\b - z_\a \sum_{i \in \b} c_i \<i|P_\b|X] \ne 0\,,
\ee
and that the angle spinor $|\hat P_\beta(z_\alpha)\>$ is defined by its CSW prescription (which has already been established at the NMHV level),
\begin{equation}\lab{takehatoff}
    |\hat P_\beta(z_\alpha)\>= \hat P_\beta(z_\alpha)|X]=P_\beta|X]=|P_\beta\>\,.
\end{equation}
The dependence on the shift \reef{commindexshift} cancels in \reef{takehatoff}. It was crucial to use the same reference spinor $|X]$ in the shift \reef{commindexshift} and in the NMHV generating function \reef{cfNMHV00} to achieve this simplification.
\begin{figure}
\begin{center}
 \includegraphics[height=3.2cm]{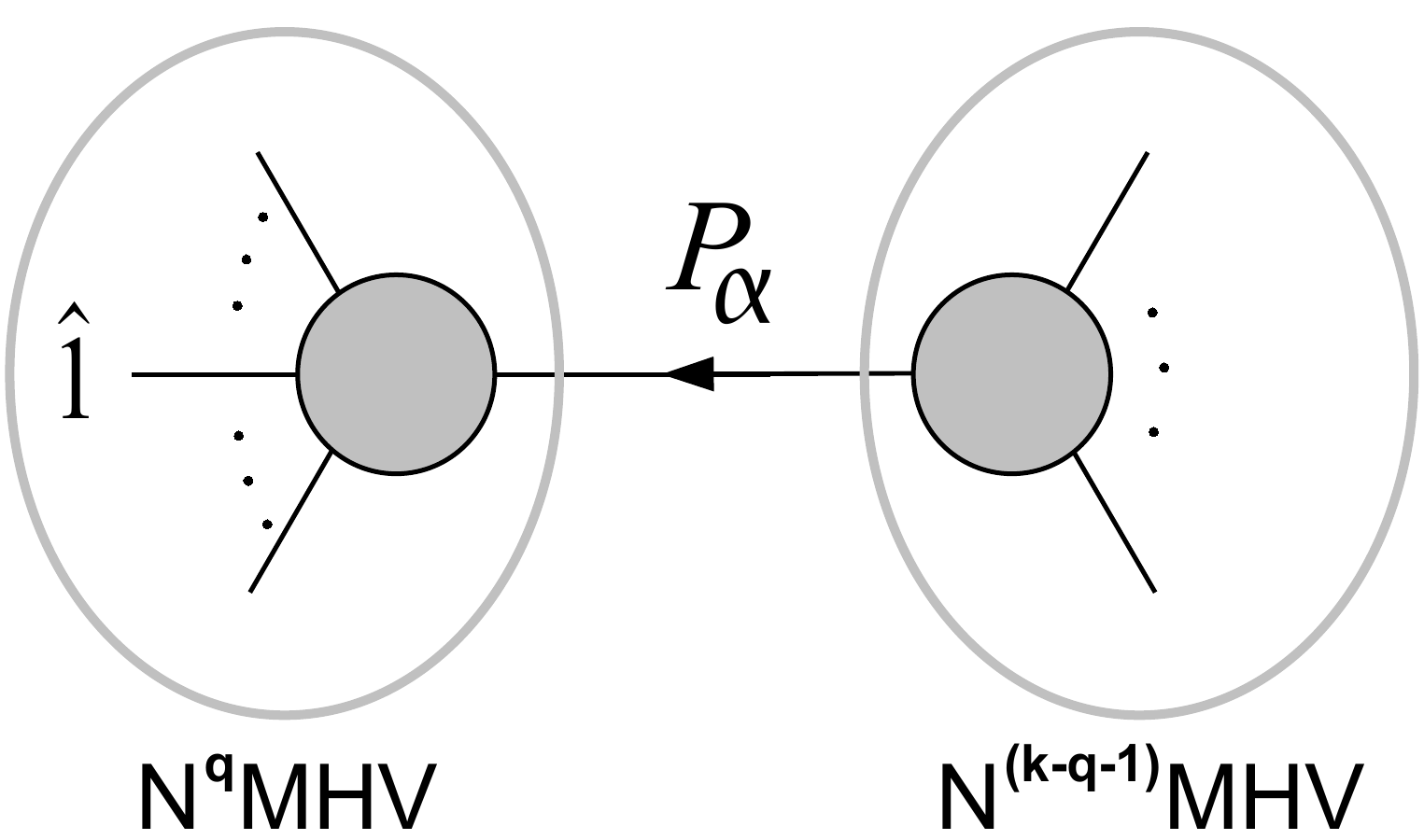}\hskip2cm \includegraphics[height=3.2cm]{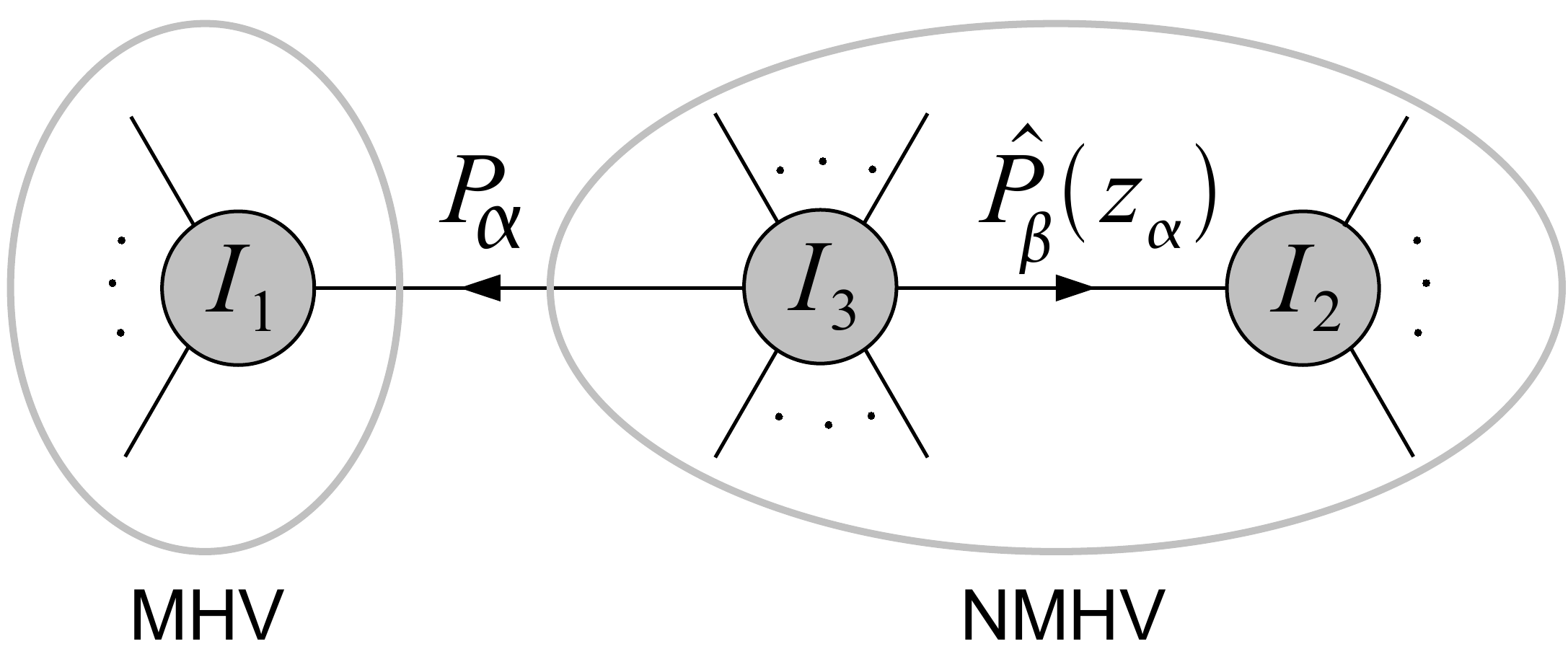}\\[3mm]
\end{center}
\vskip -.7mm
\centerline{\hskip.7cm(a)\hskip7.6cm(b)\hskip2.5cm}
\caption{(a) A diagram in the expansion~(\ref{rewritecommonindexRR}) of an N$^k$MHV amplitude under the all-line shift. ~~(b)~A~diagram in the expansion of an N$^2$MHV amplitude under the all-line shift.}
\lab{figMHVVD2shift}
\end{figure}
Thus the  shift of the external lines
only affects $\cf_\beta^{\rm NMHV}$ through the propagator $1/\hat{P}^2_\beta(z_\alpha)$, and we can therefore
write
\begin{equation}\label{cfbeta}
    \cf_\beta^{\rm NMHV}(\hat{\bar\alpha},\hat P_\alpha;\eta_{\alpha a})=\frac{P^2_\beta}{\hat P^2_\beta(z_\alpha)}\,
    \cf_\beta^{\rm NMHV}(\bar\alpha,\hat P_\alpha;\eta_{\alpha a})\,.
\end{equation}

The MHV and NMHV generating functions in~(\ref{cfMHValpha}) and~(\ref{cfbeta}) still depend on the shifted momentum $\hat P_\alpha$, but only through the angle spinor $|\hat P_\alpha\>$.  However, the dependence on $|\hat P_\alpha\>$ is homogeneous in the combined expression~(\ref{AN2MHV}). To see this, note that the derivative $D^{(4)}_\alpha$ in \reef{AN2MHV} produces four factors of $|\hat P_\alpha\>$ in the numerator while  the cyclic products $\cyc  (\hat I_1)$ and $\cyc  (\hat I_3)$  each contain two powers of $|\hat P_\alpha\>$ in the denominator.
Therefore the square
 brackets $[\hat P_\a\,X]$ from \reef{deriveCSW} cancel
in (\ref{AN2MHV}), so that we can use the CSW prescription
\begin{equation}\label{CSWpresc}
    |\hat P_\alpha\>\to |P_\alpha\>\equiv P_\alpha|X]\,.
\end{equation}
We thus obtain
\begin{equation}
    \A^{\rm N^2MHV}_{n}(1,\ldots,n)
    =D^{(16)}\sum_{\alpha}D^{(4)}_\alpha \sum_{\beta} \frac{P^2_\beta}{\hat P^2_\beta(z_\alpha)}\frac{\cf^{\rm MHV}(\alpha,- P_\alpha;\eta_{\alpha a})\,
    \cf^{\rm NMHV}_{\beta}(\bar\alpha, P_\alpha;\eta_{\alpha a})}{P^2_\alpha}\,,
\end{equation}
where the CSW prescription is understood. For each channel $\a$, we are instructed to sum over all channels $\b$ of the NMHV subamplitude, \ie~we must sum over all ordered disjoint subsets $\a$ and $\b$ of the external states. We write this sum as $\sum_{\alpha,\beta}$.

The next step is to substitute the form of the generating functions and carry out the differentiations $D^{(4)}_\alpha$. Using the $\delta$-function identity $\d^{(8)}(I)\, \d^{(8)}(J) = \d^{(8)}(I+J)\,\d^{(8)}(J)$ we obtain
\begin{equation}
\begin{split}
    &\A^{\rm N^2MHV}_{n}(1,\ldots,n)\\
    &=D^{(16)}\sum_{\alpha,\beta}D^{(4)}_\alpha
    \frac{\delta^{(8)}\bigl(\sum_{i\in{\rm ext}}|i\>\eta_{ia}\bigr)\,\delta^{(8)}\bigl(\sum_{i\in\alpha}|i\>\eta_{ia}-|P_\alpha\>\eta_{\alpha a}\bigr)}
    {\cyc  ( I_1)\cyc  ( I_2)\cyc ( I_3)~P^2_\alpha \, \hat P^2_\beta(z_\alpha)}\,\prod_{a=1}^4\sum_{i\in\beta}\<i\,P_\beta\>\eta_{ia}\\
    &=D^{(16)}\sum_{\alpha,\beta}
    \frac{\delta^{(8)}\bigl(\sum_{i\in{\rm ext}}|i\>\eta_{ia}\bigr)}
    {\cyc ( I_1)\cyc  ( I_2)\cyc ( I_3)}
    \,\frac{1}{P^2_\alpha  \,  \hat P^2_\beta(z_\alpha)}
    \,\Biggl[\,\prod_{a=1}^4\sum_{i\in\alpha}\<i\,P_\alpha\>\eta_{ia}\Biggr]\Biggl[\,\prod_{a=1}^4\sum_{i\in\beta}\<i\,P_\beta\>\eta_{ia}\Biggr]\,.
\end{split}
\end{equation}
Symmetrizing the sum over the disjoint subsets $\alpha, \beta$ gives
\begin{equation}
\lab{symab}
\begin{split}
   &\A^{\rm N^2MHV}_{n}(1,\ldots,n)\\
   &=D^{(16)}\frac{1}{2}\sum_{\alpha,\beta}\!
    \frac{\delta^{(8)}\bigl(\sum_{i\in{\rm ext}}|i\>\eta_{ia}\bigr)}
    {\cyc  ( I_1)\cyc  ( I_2)\cyc ( I_3)}
    \Biggl[\!\frac{1}{P^2_\alpha  \,  \hat P^2_\beta(z_\alpha)}+\frac{1}{ \hat P^2_\alpha(z_\beta)  \, P^2_\beta}\Biggr]\!
    \Biggl[\,\prod_{a=1}^4\sum_{i\in\alpha}\<i\,P_\alpha\>\eta_{ia}\!\Biggr]\!\Biggl[\,\prod_{a=1}^4\sum_{i\in\beta}\<i\,P_\beta\>\eta_{ia}\!\Biggr]\,.
\end{split}
\end{equation}
Each term in the sum $\sum_{\alpha,\beta}$ is an MHV vertex diagram, as the one shown in Fig.~\ref{figMHVVD2shift}b. The symmetrization in $\a, \b$ is equivalent to having obtained the same MHV vertex diagrams by expanding the amplitude as NMHV $\times$ MHV instead of MHV $\times$ NMHV. Thus the symmetrization
counts each MHV vertex diagram twice, and we have compensated with a factor of $1/2$ in \reef{symab}.

Next we use the identity
\begin{equation}
    \frac{1}{P^2_\alpha  \hat P^2_\beta(z_\alpha)}+\frac{1}{ \hat P^2_\alpha(z_\beta) P^2_\beta}=\frac{1}{P^2_\alpha P^2_\beta}
\end{equation}
which follows from the contour integral identity  \cite{greatdane}
\begin{equation}\label{contint}
    0\,=\,\frac{1}{2\pi i}\oint \frac{dz}{z} \frac{1}{ \hat P^2_\a(z)}\frac{1} { \hat P^2_\b(z)}
    \,=\,\frac{1}{P^2_\alpha P^2_\beta}-\frac{1}{P^2_\alpha \hat P^2_\beta(z_\alpha)}-\frac{1}{ \hat P^2_\alpha(z_\beta) P^2_\beta}
    \,,
\end{equation}
where the contour  encircles all poles, and
\begin{equation}
  \hat P_\a^2(z)  = P_\a^2 - z \sum_{i\in \a}c_i \<i|P_\a|X]\,,\qquad
  \hat P_\b^2(z)  = P_\b^2 - z \sum_{i\in \b} c_i\<i|P_\b|X]\,.
\end{equation}

Finally we  rename $\alpha\to\alpha_1$ and $\beta\to\alpha_2$
to conclude
\begin{equation}
    \A^{\rm N^2MHV}_{n}(1,\ldots,n)
    =D^{(16)}\Biggl\{\frac{1}{2}
    \sum_{\alpha_1,\alpha_2}
    \frac{\delta^{(8)}\bigl(\sum_{i\in{\rm ext}}|i\>\eta_{ia}\bigr)}
    {\cyc  ( I_1)\cyc  ( I_2)\cyc ( I_3)}
    \prod_{A=1}^2
    \Biggl[\,\frac{1}{P^2_{\alpha_A}}
    \prod_{a=1}^4\sum_{i\in\alpha_A}\<i\,P_{\alpha_A}\>\eta_{ia}\Biggr]\Biggr\}\,.
\end{equation}
Note that all effects of the shift
 have
canceled in the final formula.
 The sum over $\alpha_1,\alpha_2$ is a sum over MHV vertex diagrams. Each MHV vertex diagram appears twice, due to the symmetry $\alpha_1\leftrightarrow\alpha_2$ in the summation. We conclude that
\begin{equation}\label{FnN2MHV}
    \cf^{\rm N^2MHV}_n(\eta_{1a},\ldots,\eta_{na})
    =\hskip-.2cm\sum_{\stackrel{\text{MHV vertex diagrams }}{\{\alpha_1,\alpha_2\}}}\hskip-.15cm
    \frac{\delta^{(8)}\bigl(\sum_{i\in{\rm ext}}|i\>\eta_{ia}\bigr)}
    {\,\cyc  ( I_1)\cyc  ( I_2)\cyc ( I_3)}
    \prod_{A=1}^2
    \Biggl[\,\frac{1}{P^2_{\alpha_A}}
    \prod_{a=1}^4\sum_{i\in\alpha_A}\<i\,P_{\alpha_A}\>\eta_{ia}\Biggr]\,.
\end{equation}
By comparison with~(\ref{FnNkMHV}) we see that this is exactly the  desired N$^2$MHV generating function.   The MHV vertex expansion of  any amplitude at this level is obtained by
applying the appropriate Grassmann differential operator $D^{(16)}$.
This concludes the proof of the MHV vertex expansion at the N$^2$MHV level.
Next we extend our argument to level N$^k$MHV, for $k \ge 3$.

\subsection{The MHV vertex expansion for all $\cn=4$ SYM amplitudes}\label{secMHVVEall}
To generalize our result to N$^k$MHV amplitudes for all $k$, we proceed inductively in $k$ and assume that the MHV vertex expansion is valid for
all $q<k$. We can then use the generating function~(\ref{FnNkMHV}) for both subamplitudes in the expansion~(\ref{rewritecommonindexRR}) of an N$^k$MHV amplitude under an all-line shift.  As indicated in Fig.~\ref{figxyx}
\begin{figure}
\begin{center}
 \includegraphics[height=4.5cm]{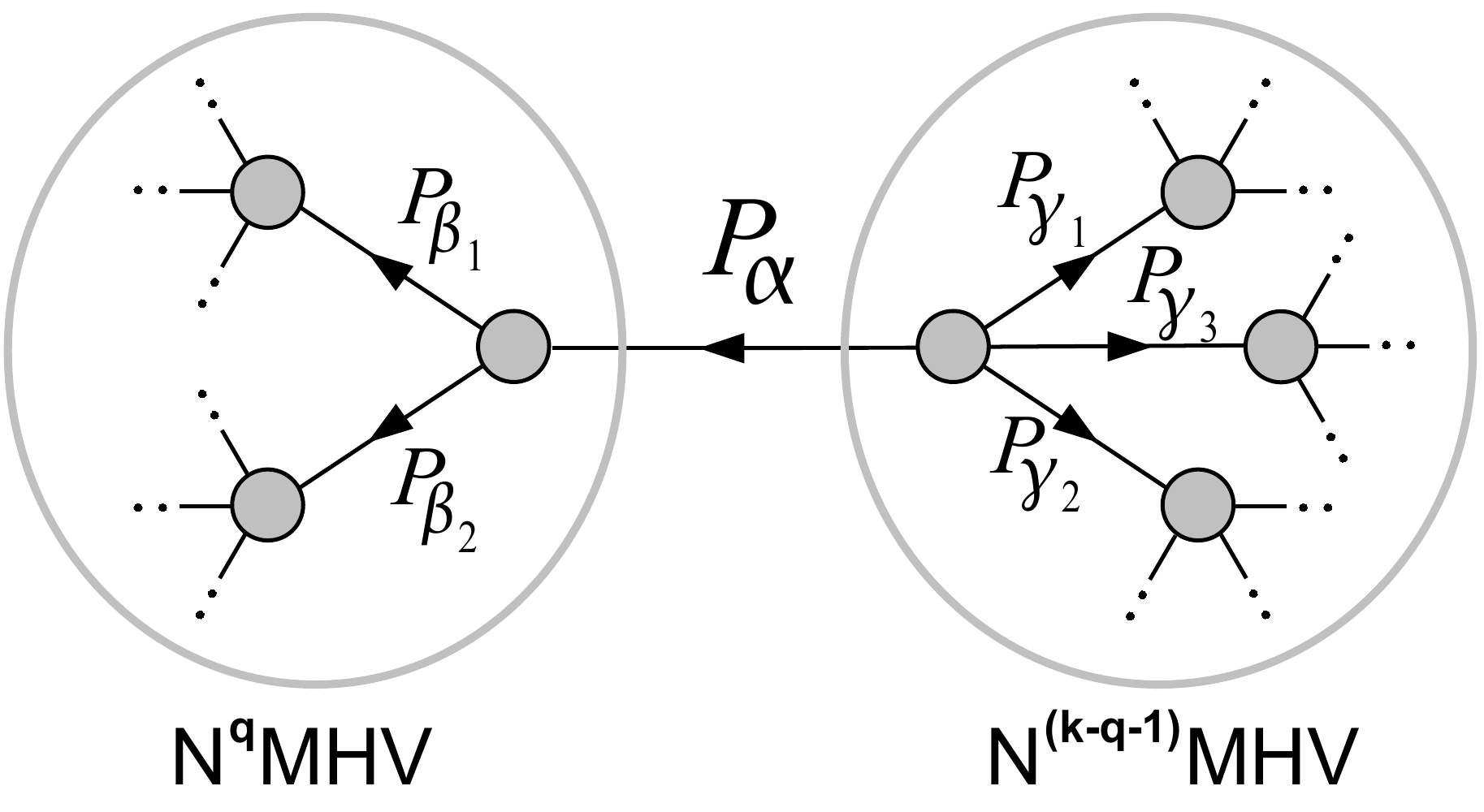}\\[3mm]
\end{center}
\vskip -.7mm
\caption{The MHV vertex expansion is applied to the subamplitudes in the all-line recursion relation. As the arrows indicate, channels $\b_A$ and
$\g_B$ are chosen to include external states only. The dots in the figure represent the remaining parts of the MHV vertex diagrams.}
\lab{figxyx}
\end{figure}
 the generating function $\cf^\text{N$^q$MHV}$
contains a sum over channels $\b_A,~ A=1,\dots,q\,$, and $\cf^\text{N$^{(k-q-1)}$MHV}$ contains a sum over
channels $\g_B,~ B=1,\ldots, k-q-1.$  Channels have been chosen to include only external states
rather than the
 internal line
$\pm \hat{P}_\a$.  In more detail, the generating function  $\cf^\text{N$^q$MHV}$
is given by
\bea
 \nonumber
 \cf^\text{N$^{q}$MHV}(\hat\alpha,-\hat P_\alpha;\eta_{\alpha a})
 &=& \!\!\!\!\!\!\! \sum_{\stackrel{\text{MHV vertex diagrams}}{\{\beta_1,\ldots,\beta_q\}}}\!\!\!\!\!\!\!
 \frac{ \d^{(8)}\bigl(\sum_{i\in \alpha}|i\>\h_{ia}\!-\!|\hat P_\alpha\>\eta_{\alpha a}\bigr)}{~~\cyc (\hat I_1)~\cdots\cyc (\hat I_{q+1})}
 \prod_{A=1}^q\Biggl[\frac{1}{\hat P_{\beta_{\!A}}^2\!(z_\alpha)}\prod_{a=1}^4\sum_{i\in \beta_A}\!\<i\,\hat P_{\beta_{\!A}}\!(z_\alpha)\>\h_{ia}\!\Biggr]\\
 &\equiv& \!\!\!\!\!\!\! \sum_{\stackrel{\text{MHV vertex diagrams}}{\{\beta_1,\ldots,\beta_q\}}}\!\!\!
 \cf^\text{N$^{q}$MHV}_{\beta_1,\ldots,\beta_q}(\hat\alpha,-\hat P_\alpha;\eta_{\alpha a})\,.
\lab{leftgen}
\eea
The momenta $\hat{P}_{\b_A}(z_a)$ are shifted as in \reef{pbsq}, and we choose the same reference spinor $|X]$ so that \reef{takehatoff} holds for all
 spinors $|\hat P_{\beta_A}(z_\alpha)\>$.
The effect of the shift is now confined to  the
 angle spinor
 $|-\hat{P}_\a\>$, which is given
by \reef{deriveCSW}, and the propagators $1/P^2_{\b_A}(z_\a)$.  For each configuration
of channels $\{\b_1,\ldots,\b_q\}$
in    \reef{leftgen} we follow \reef{cfbeta} and obtain
\begin{equation}
    \cf^\text{N$^{q}$MHV}_{\beta_1,\ldots,\beta_q}(\hat\alpha,-\hat P_\alpha;\eta_{\alpha a})
    =\frac{P_{\beta_1}^2\cdots P_{\beta_q}^2}{\hat P_{\beta_1}^2(z_\alpha)\cdots \hat P_{\beta_q}^2(z_\alpha)}\,\cf^\text{N$^{q}$MHV}_{\beta_1,\ldots,\beta_q}(\alpha,-\hat P_\alpha;\eta_{\alpha a})\,.
\end{equation}

We treat the  generating function  $\cf^\text{N$^{(k-q-1)}$MHV}$ in the same way, but note that the $\d^{(8)}$
in its numerator reads
 $\d^{(8)}(\sum_{i\in\bar{\a}} |i\>\h_{ia} +|\hat{P}_\a \>\h_{\a a})$.
For each
configuration of channels $\gamma_1,\ldots,\gamma_{k-q-1}$ we write
\begin{equation}
\begin{split}
 \cf^\text{N$^{(k-q-1)}$MHV}(\hat{\bar\alpha},\hat P_\alpha;\eta_{\alpha a})
 &= \!\!\! \sum_{\stackrel{\text{MHV vertex diagrams }}{\{\gamma_1,\ldots,\gamma_{k-q-1}\}}}\!\!\!
 \cf^\text{N$^{(k-q-1)}$MHV}_{\gamma_1,\ldots,\gamma_{k-q-1}}(\hat{\bar\alpha},\hat P_\alpha;\eta_{\alpha a})
\end{split}
\end{equation}
with
\begin{equation}
    \cf^\text{N$^{(k-q-1)}$MHV}_{\gamma_1,\ldots,\gamma_{k-q-1}}(\hat{\bar\alpha},\hat P_\alpha;\eta_{\alpha a})
    =\frac{P_{\gamma_1}^2\cdots P_{\gamma_{k-q-1}}^2}{\hat P_{\gamma_1}^2(z_\alpha)\cdots \hat P_{\gamma_{k-q-1}}^2(z_\alpha)}
    \cf^\text{N$^{(k-q-1)}$MHV}_{\gamma_1,\ldots,\gamma_{k-q-1}}({\bar\alpha},\hat P_\alpha;\eta_{\alpha a})\,.
\end{equation}
The product of generating functions in~(\ref{rewritecommonindexRR}) is again homogeneous in the angle spinor $|\hat P_\alpha\>$.  Again this allows us to replace
$|\hat P_\alpha\>$
by the CSW
spinor $|P_\a\> = P_\a|X]$.  We then have
\begin{equation}
\begin{split}
    &D^{(4k+8)}D^{(4)}_\alpha\Biggl[\frac{\cf^\text{N$^{q}$MHV}_{\beta_1,\ldots,\beta_q}(\hat\alpha,-\hat P_\alpha;\eta_{\alpha a})\,
    \cf^\text{N$^{(k-q-1)}$MHV}_{\gamma_1,\ldots,\gamma_{k-q-1}}(\hat{\bar\alpha},\hat P_\alpha;\eta_{\alpha a})}{P_\alpha^2}\Biggr]\\
 &= D^{(4k+8)}\frac{\d^{(8)}\bigl(\sum_{i\in {\rm ext}}|i\>\h_{ia}\bigr)}{\cyc (I_1)\cdots\cyc (I_{k+1})} ~
 \Biggl[\frac{1}{ P_{\alpha}^2}\prod_{a=1}^4\sum_{i\in \alpha}\<i\, P_{\alpha}\>\h_{ia}\Biggr]
 \prod_{A=1}^q\,\Biggl[\frac{1}{\hat P_{\beta_A}^2\!(z_\alpha)}\prod_{a=1}^4\sum_{i\in \beta_A}
 \<i\,P_{\beta_A}\>\h_{ia}\Biggr] \\
 &\hskip7.7cm\times  \prod_{B=1}^{k-q-1}\Biggl[\frac{1}{\hat P_{\gamma_B}^2\!(z_\alpha)}\prod_{a=1}^4\sum_{i\in \gamma_B}
 \<i\,P_{\gamma_B}\>\h_{ia}\Biggr]\,.
\end{split}
\end{equation}
Every set $\{\alpha,\beta_1,\ldots,\beta_q,\gamma_1,\ldots,\gamma_{k-q-1}\}$ represents an MHV vertex diagram of the full N$^k$MHV amplitude. In fact, summing over channels $\beta_A$ and $\gamma_B$
we obtain every possible MHV vertex diagram containing the internal line $\alpha$ exactly once.
We need to express the result as  sum over all MHV vertex diagrams and note that
  any fixed MHV vertex diagram, with channels $\{\alpha_1,\ldots,\alpha_k\},$ is contained in the sum over $\alpha$, $\beta_A$, and $\gamma_A$ exactly $2k$ times, namely once with $\alpha=\alpha_B$ and once with $\alpha=\bar \alpha_B$, for each $B=1,\ldots,k$. We conclude that
\begin{equation}
\begin{split}
 &\A^\text{N$^k$MHV}_{n}(1,\ldots,n)\\
 &=D^{(4k+8)}\frac{1}{2}\sum_\alpha
  \sum_{\stackrel{\text{MHV vertex diagrams }}{\{\beta_1,\ldots,\beta_q\} ,\{\gamma_1,\ldots,\gamma_{k-q-1}\}}}\hskip-.5cm
    D^{(4)}_\alpha\Biggl[\frac{\cf^\text{N$^{q}$MHV}_{\beta_1,\ldots,\beta_q}(\hat\alpha,-\hat P_\alpha;\eta_{\alpha a})\,
    \cf^\text{N$^{(k-q-1)}$MHV}_{\gamma_1,\ldots,\gamma_{k-q-1}}(\hat{\bar\alpha},\hat P_\alpha;\eta_{\alpha a})}{P_\alpha^2}\Biggr]\\
 &=D^{(4k+8)}\hskip-.5cm \sum_{\stackrel{\text{MHV vertex diagrams }}{\{\alpha_1,\ldots,\alpha_k\}}}\sum_{B=1}^k
 \frac{ \d^{(8)}\Big(\sum_{i\in{\rm ext}}|i\>\h_{ia}\Bigr)\prod_{A=1}^k\Bigl[\prod_{a=1}^4\sum_{i\in \alpha_A}\<i\,P_{\alpha_A}\>\h_{ia}\Bigr]}
 {\cyc(I_1)\cdots\cyc (I_{k+1})\times
 \hat P^2_{\alpha_1}(z_{\alpha_B}\!)\cdots \hat P^2_{\alpha_{\!B-1}}\!(z_{\alpha_B}\!) \, P^2_{\alpha_{\!B}}\,\hat P^2_{\alpha_{\!B+1}}\!(z_{\alpha_B}\!)\cdots \hat P^2_{\alpha_k}\!(z_{\alpha_B}\!)}\,.
\end{split}
\end{equation}
Next, we use  the identity
\begin{equation}
    \sum_{B=1}^k
    \frac{1}{\hat P^2_{\alpha_1}(z_{\alpha_B})\cdots \hat P^2_{\alpha_{\!B-1}}\!(z_{\alpha_B}) \, P^2_{\alpha_{\!B}}\,\hat P^2_{\alpha_{\!B+1}}\!
    (z_{\alpha_B})\cdots \hat P^2_{\alpha_k}\!(z_{\alpha_B})}
    =\frac{1}{P_{\alpha_1}^2\cdots P_{\alpha_k}^2}\,,
\end{equation}
which follows from a contour integral similar to \reef{contint}.  Finally, we obtain
\begin{equation}
\begin{split}
 &\A^\text{N$^k$MHV}_{n}(1,\ldots,n)
 =D^{(4k+8)}\sum_{\stackrel{\text{MHV vertex diagrams }}{\{\alpha_1,\ldots,\alpha_k\}}}
 \frac{ \d^{(8)}\bigl(\sum_{i\in{\rm ext}}|i\>\h_{ia}\bigr)}{\cyc (I_1)\cdots\cyc (I_{k+1})} ~
 \prod_{A=1}^k\Biggl[\frac{1}{P_{\alpha_A}^2}\prod_{a=1}^4\sum_{i\in \alpha_A}\<i\,P_{\alpha_A}\>\h_{ia}\Biggr]\,.
\end{split}
\end{equation}
The differential operator $D^{(4k+8)}$ associated with the amplitude thus acts precisely on the generating function~(\ref{FnNkMHV}) associated with the MHV vertex expansion.
This concludes  the inductive derivation of the MHV vertex expansion for all $k$. In the derivation we used recursion relations resulting from all-line shifts \reef{commindexshift}.
The purpose of the following section is to prove the validity of all-line recursion relations.

\setcounter{equation}{0}
\section{Large $z$ behavior under all-line shifts}\label{secvalidity}

In the previous section we established the MHV vertex expansion using all-line shift recursion relations.
To prove its validity, we must show that  $A^\text{N$^k$MHV}_n(z) \to 0$ as  $z \to \infty$ under any all-line shift when $k \ge 1$. In this section we prove a stronger result, namely
\be
\lab{2Bproven}
\boxed{~\phantom{\Bigg(}
 \A^\text{N$^k$MHV}_n(z) \sim O(z^{-k})
 ~~~~~\text{as}~~~~~z \to \infty~~~~~
 \text{under any all-line shift, ~for all}
 ~~~~k \ge -1\,.
~~}
\ee

Let us first consider the two simplest cases, MHV and anti-MHV:
\begin{itemize}
\item {\bf MHV:} For $k=0$, the statement \reef{2Bproven} becomes
$\A^\text{MHV}_n(z) \sim O(1)$. This is clearly true, since MHV amplitudes depend only on angle brackets which are left untouched by the square spinor shift.

\item {\bf anti-MHV:}
At the $k$th next-to-MHV level, amplitudes with $n=k+4$ external legs are anti-MHV. The conjugate Parke-Taylor amplitude for $\A^\text{N$^k$MHV}_{k+4} = \A^{\overline{\text{MHV}}}_{k+4}$ involves 4 square brackets in the numerator and $k+4$ square brackets in the denominator. Under an all-line shift, every square bracket goes as $O(z)$ for large $z$, so the amplitude behaves as $O(z^{-k})$. This proves \reef{2Bproven} for all anti-MHV amplitudes with $k\ge-1$.

\end{itemize}

For all other amplitudes,
 we prove the validity of \reef{2Bproven}
by an inductive argument whose starting points are the
 above
MHV and anti-MHV results. Our strategy is to express $\A^\text{N$^k$MHV}_n$ in terms of a valid BCFW recursion relation which consists of diagrams involving subamplitudes $\A^\text{N$^{k'}$MHV}_{n'}$ with $k' \le k$ and $n'<n$. If \reef{2Bproven} holds for these lower-point amplitudes, then each diagram of the BCFW
 expansion
can be shown
to be $O(z^{-k})$
 under the all-line shift.

\subsection{BCFW representation of N$^k$MHV amplitudes}
\lab{s:bcfw}

Every N$^k$MHV tree amplitude
 of $\cn =4$ SYM with $k \ge 1$
admits at least one valid BCFW 2-line shift, and it can therefore be expressed in terms of a BCFW recursion relation. We proved this in \cite{efk1}.

Consider an N$^k$MHV amplitude, $k \ge 1$,  with external lines $1,\dots,n$ labeled such that a BCFW $[1,\ell\>$-line shift gives a valid recursion relation for the amplitude. Details of the shift are presented in appendix \ref{app:shifts}. In the BCFW representation, the amplitude is a sum
 of
diagrams of the type
\bea
  \lab{bcfwdiag}
  \A^\text{N$^{k_1}$MHV}_{n_1}(I) ~ \times ~ \frac{1}{P_I^2}~
  \times ~ \A^\text{N$^{k_2}$MHV}_{n_2}(J)
\eea
with
\bea
  \lab{n12k12}
  n_1 + n_2 = n+2
   \,,\quad n_{1},n_2\geq 3\,,
  ~~~~~\text{and}~~~~~
  k_1 + k_2 = k -1
  \,,\quad k_{1} \ge 0,\quad k_2\geq -1
  \,.
\eea
As depicted in figure~\ref{figBCFW},
subamplitudes $I$ and $J$ are chosen such that $I$ contains the line $\tilde{1}$ and $J$ contains $\tilde{l}$. (We use tilde's to
denote the BCFW shifted momenta.)
Then
\bea
  \A^\text{N$^{k_1}$MHV}_{n_1}(I)
  ~=~\A^\text{N$^{k_1}$MHV}_{n_1}(\tilde{1},\dots,-\tilde{P}_I,\dots) \, ,
  \hspace{0.7cm}
  \A^\text{N$^{k_2}$MHV}_{n_2}(J)
  ~=~\A^\text{N$^{k_2}$MHV}_{n_2}(\tilde{\ell},\dots,\tilde{P}_I,\dots) \, .
\eea

\begin{figure}
\begin{center}
 \includegraphics[height=2cm]{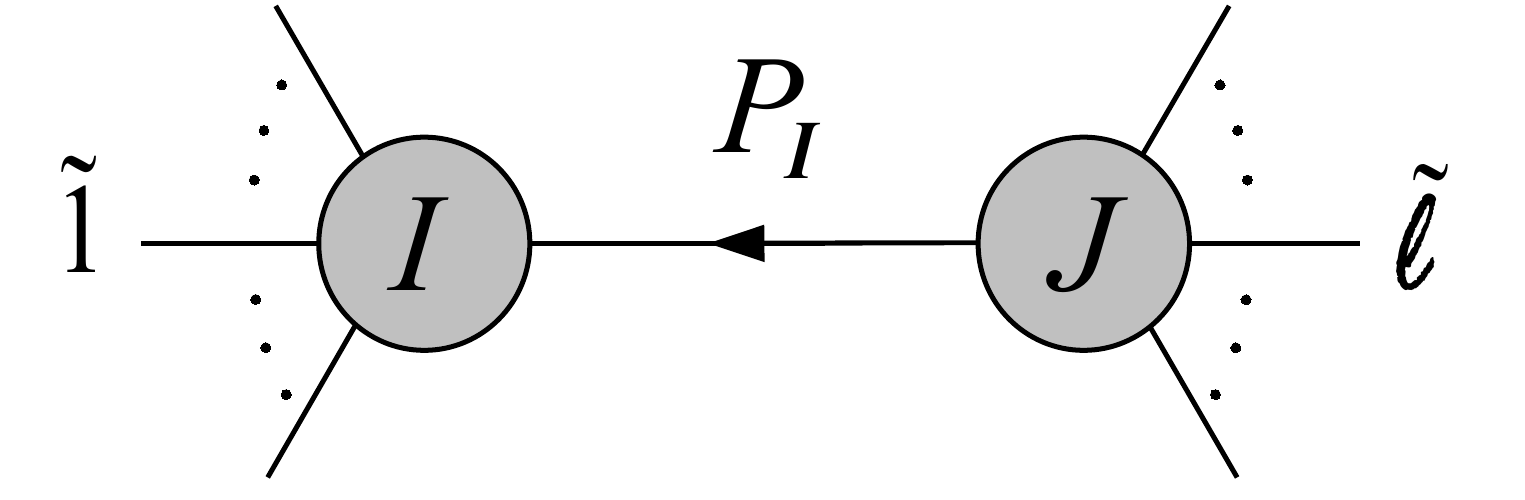}\\[3mm]
\end{center}
\vskip -.7mm
\caption{A diagram in the BCFW recursion relation associated with a $[1,\ell\>$-shift.}
\lab{figBCFW}
\end{figure}

An argument identical to the one around eq.~\reef{bsp} shows that subamplitude $I$ vanishes for $k_1 = -1$, \ie if it is a 3-point anti-MHV vertex. The conjugate statement applies to subamplitude $J$ which vanishes if it is a 3-point MHV vertex. Hence, whenever $J$ is a 3-point vertex it must be anti-MHV,
\bea\label{extraconstraints}
  n_2 = 3  ~~\implies~~ k_2 = -1\, .
\eea
This will be important in our analysis.


\subsection{Kinematics of the shifts}
\lab{s:kin}

In our proof, we perform the all-line
 shift
\reef{commindexshift} on the BCFW representation of the amplitude $\A^\text{N$^k$MHV}_n$.
Details of how the
 all-line shift
acts on BCFW shifted lines
$\tilde{1}$, $\tilde{\ell}$ and $\tilde{P}_I$ are given in appendix \ref{app:shifts} and we simply summarize the needed results here.

 As in \reef{commindexshift}, we
let hats denote momenta shifted under the
 all-line shift.
For large $z$, we find
\bea
   \frac{1}{\hat P_I^2} ~\sim~ O(z^{-1})\, ,~~~~~~
   |\hat{\tilde{P}}_{I}\> ~\sim~
   O(1) \, ,~~~~~~
   | \hat{\tilde{P}}_I ] ~=~ z
   \!\sum_{{\rm ext}\, i\in I}\!
   c_i\frac{\<1i\>}{\<1\ell\>}|X] + O(1) \, .
\eea
The subset conditions \reef{subsetsumci} are necessary to ensure that $|\hat{\tilde{P}}]$ has the needed $O(z)$ shift.
The all-line shift also affects the BCFW-shifted lines
 $\tilde{1}$ and $\tilde{\ell}$.
For large $z$, we have
$|\hat{\tilde{1}}]  \sim O(z)$ and $|\hat{\tilde{\ell}}\> \sim O(1)$.

Under the
all-line shift,
all spinor products involving any combination of the square spinors $| \hat{\tilde{P}}_I ]$, $|\hat{\tilde{1}}]$, $|\hat{\ell}]$, and other  external lines $|\hat i]$ go as $O(z)$. Furthermore, all spinor products involving any combination of  the angle spinors $| \hat{\tilde{P}}_I \>$, $|\hat{\tilde{\ell}}\>$, $|\hat{1}\>$, and other  external lines $|i\>$ go as $O(1)$, \emph{except} when subamplitude $J$ has $n_2=3$ lines. In this exceptional case, the angle spinors of this subamplitude are $O(1/z)$.  However, this plays no role since \reef{extraconstraints} ensures that the vertex $J$ is
 then
anti-MHV and therefore only depends on square brackets.
We conclude that:
\begin{quote}
\emph{At large $z$, an all-line shift on the whole amplitude effectively
acts as an all-line shift on the subamplitudes in the BCFW representation.}
\end{quote}

The large $z$ behavior summarized above holds if the choice of complex parameters $c_i$ is \emph{generic}, \ie~if the coefficients do not satisfy any accidental relations which affect the large $z$ behavior.
For instance, one must ensure that the subset conditions \reef{subsetsumci} are satisfied on the effective all-line shifts on both subamplitudes. Again, we refer to appendix~\ref{app:shifts} for details.


\subsection{Proof that $\A^\text{N$^k$MHV}_n(z) \sim z^{-k}$
as $z \to \infty$ under all-line shifts}
\lab{s:ind}
\begin{figure}
\centerline{
\includegraphics[width=5.5cm]{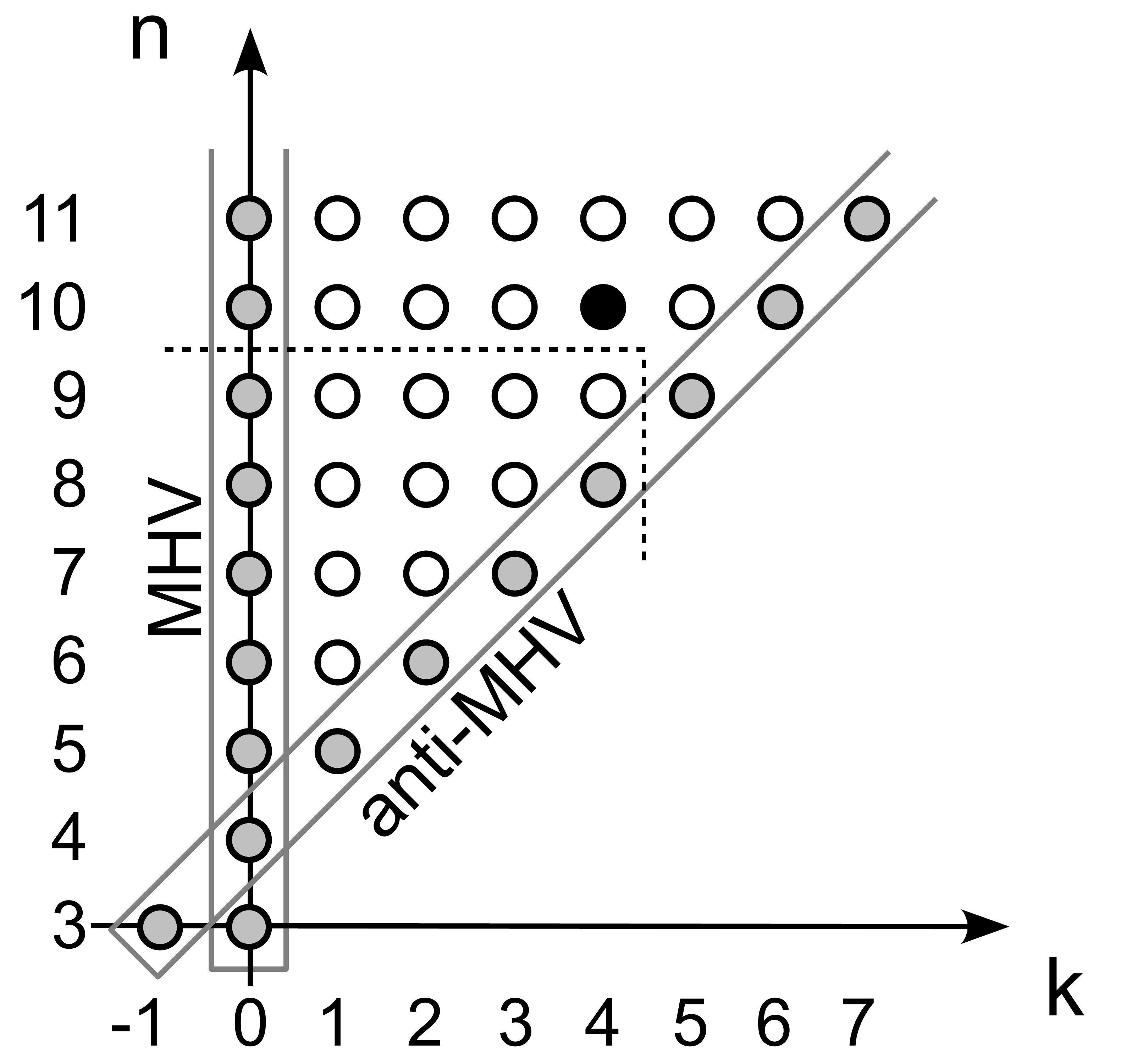}}
\caption{N$^k$MHV$_n$ amplitudes represented as dots
 in a triangular region of
the $(k,n)$-plane.
The $O(z^{-k})$ falloff under the all-line shift is directly established for MHV and anti-MHV amplitudes, which are represented by the gray dots on the boundary of the region.
The BCFW expansion of an
 N$^k$MHV$_n$ amplitude involves \emph{lower-point}
 subamplitudes at level N$^{k'}$MHV with $k' \le k$.
 For example,
 amplitudes at the black
  dot
 depend only on subamplitudes
  represented by dots
 to the left and below the dashed lines.
Starting the induction at $(k,n)=(1,6)$, one can progressively move up in the interior of the triangular region and thus prove the $O(z^{-k})$-falloff for all amplitudes.}
\lab{fig:induct}
\end{figure}

Consider any amplitude $\A^\text{N$^k$MHV}_n$ with $k\ge 1$ and $n \ge 6$. As our \emph{inductive assumption} we assume that all N$^{k'}$MHV$_{n'}$ amplitudes with
$-1\le k'\le k$ and $n'<n$
go as $O(z^{-k'})$ at
 large $z$ under
 any all-line shift.

Let $\A^\text{N$^k$MHV}_n$ be represented by the valid BCFW recursion relation discussed in section \ref{s:bcfw}. The diagrams of this representation are of the form \reef{bcfwdiag}.
Perform now the all-line shift on the amplitude $\A^\text{N$^k$MHV}_n$. As summarized in section \ref{s:kin}, this shift acts on the subamplitudes $I$ and $J$ as an all-line shift. Since $n_{1,2} < n$ and $k_{1,2} \le k$ our inductive assumption guarantees that
 $\A^\text{N$^{k_1}$MHV}_{n_1}(\hat I)\sim z^{-k_1}$ and $\A^\text{N$^{k_2}$MHV}_{n_2}(\hat J)\sim z^{-k_2}$
for large $z$.
The internal momentum also shifts, so the propagator contributes an extra order of suppression, $1/z$. We
therefore conclude that the BCFW diagram falls off as
\bea
  \A^\text{N$^{k_1}$MHV}_{n_1}(\hat I)
  ~ \times ~ \frac{1}{\hat P_I^2}~
  \times ~ \A^\text{N$^{k_2}$MHV}_{n_2}(\hat J)
  ~~\sim~~ O(z^{-k_1})\, O(z^{-1})\, O(z^{-k_2})
  ~~\sim~~ O(z^{-k})\,
\eea
for large $z$. We have used that
$k_1 + k_2 +1 = k$.
We conclude that the whole amplitude $\A^\text{N$^k$MHV}_n$ falls off
 at least as $1/z^k$
as $z \to \infty$.

We must establish a basis of induction that allows us
to carry through the above inductive argument for all N$^k$MHV$_n$ amplitudes. Figure \ref{fig:induct} illustrates that the established $O(z^{-k})$ large $z$ behavior of MHV and anti-MHV amplitudes suffices to guarantee that one can recursively reach the desired conclusion for all N$^k$MHV$_n$ amplitudes via the BCFW representation.
 The large $z$ behavior of  MHV and anti-MHV amplitudes under all-line shifts thus provides the required basis of induction and completes the proof.

\setcounter{equation}{0}
\section{$\cn=4$ SYM amplitudes under square spinor shifts}\label{secsquare}
In the previous sections we have established the validity of the MHV vertex expansion for all $\cn=4$ SYM amplitudes. In this section we use the MHV vertex expansion as a tool to analyze the large $z$ behavior of amplitudes under various classes of square spinor shifts. This analysis leads to new recursion relations for $\cn=4$ SYM. In particular, we will find that all amplitudes $\A^\text{N$^k$MHV}_{n}$ permit shifts of only $k+2$ instead of all $n$ lines under which they fall off at least as fast as $1/z^k$.

In the following, we will consider square spinor shifts
of $s\geq3$ lines $m_1,\ldots,m_s$
\begin{equation}\label{generalsquare}
    |m_i] \to |m_i] + z \, c_i |Y]\,,\quad i=1,\ldots,s\,,
\end{equation}
with momentum conservation imposed:
\begin{equation}\label{sumcisquare}
    \sum_{i=1}^s c_i|m_i\>=0\,.
\end{equation}

We will study the large $z$ behavior of an amplitude $\A^\text{N$^k$MHV}_n$ under such shifts, by studying its representation under the MHV vertex expansion.
A diagram in the MHV vertex expansion of $\A^\text{N$^k$MHV}_n$ takes the form
\bea
  \lab{mhvform}
  \frac{\MHV(I_1) \cdots \MHV(I_{k+1})}{P_{\alpha_1}^2 \cdots P_{\alpha_{k}}^2}\,.
\eea
Typically each diagram depends on the reference spinor $|X]$ through the CSW prescription. However, as the sum of diagrams is independent of $|X]$, we are free to choose any $|X]$, in particular we can set $|X] = |Y]$.  This choice of reference spinor in the MHV vertex expansion implies that all
 $|P_{\alpha_A}\> = P_{\alpha_A}|Y]$
remain unshifted under \reef{generalsquare}. Since the MHV vertices depend just on angle brackets, only the propagators of the MHV vertex diagram \reef{mhvform} shift.

By inspecting the diagrams of the MHV vertex expansion, we now derive the following four results:
\begin{enumerate}
\item
{\bf Any amplitude $\A^\text{N$^k$MHV}_n(z)$ goes as $O(1)$, or better,
under \emph{any} square spinor shift
of the external lines.}

\emph{Proof:}
Consider any MHV vertex diagram  under a general square spinor shift~(\ref{generalsquare}).
The only parts of \reef{mhvform} that shift are the propagators. However, as we are not imposing any rules on which lines we shift, all shifted lines could sit on the same MHV vertex, and in that case no propagator shifts. Therefore the worst possible large $z$ behavior of an MHV vertex diagram is $O(1)$. Thus
$\A^\text{N$^k$MHV}_n(z) \sim O(1)$, or better,
under \emph{any} shift of $s\ge 3$ square spinors.

\item
{\bf Any amplitude $\A^\text{N$^k$MHV}_n(z)$ goes as $O(z^{-k})$, or better,
under a \emph{common-index shift}.}

\emph{Definition:} A common-index shift of an N$^k$MHV amplitude is a square spinor shift
of the form~(\ref{generalsquare}) with $s=k+2$ lines shifted. The particles on the shifted lines  $m_i$ are required to all carry at least one common $SU(4)$ index, say $a$. Furthermore, we require that the shift parameters $c_i$ satisfy
\begin{equation}\label{commonindexci}
    \sum_{\text{subset }m_i} c_i|m_i\>\neq 0
\end{equation}
for any  ordered\footnote{We consider a subset $m_i,\ldots, m_f$ of the shifted lines $m_1,\ldots,m_s$ as \emph{ordered}, if the set contains all \emph{shifted} external lines between $m_i$ and $m_f$. We do not require that the lines $m_i,\ldots, m_f$ are consecutive, \ie that there are no unshifted lines between them.}
proper subset of the shifted lines $\{m_1,\ldots,m_s\}$.

Any $\cn=4$ SYM amplitude permits
 a common-index shift.
In fact, a generic amplitude allows four distinct common-index shifts --- one for each choice of $SU(4)$ index $a$. For some amplitudes, different $SU(4)$ indices imply the same common-index shift. Pure gluon amplitudes, for example, only admit one unique common-index shift, namely the shift of all negative helicity lines.

\emph{Proof:}
Consider an N$^k$MHV amplitude. Each diagram of its MHV vertex expansion contains $k$ internal lines and $k+1$ MHV vertices. We distinguish between ``end-vertices" with a single internal line and ``middle-vertices" with two or more internal lines.
We perform a common-inde shift as explained above, with  the $k+2$ shifted lines sharing at least one common index $a$.
The $k+1$ MHV vertices must each have precisely 2 lines carrying the index $a$. Each end-vertex must contain either 1 or 2 of the external shifted lines $m_i$, since the single internal line can supply at most one index $a$. Thus the momentum carried by each internal line in the diagram contains at least one, but not all shifted momenta.
It then follows from~(\ref{commonindexci}) that \emph{all} $k$ internal lines of the diagram must shift.
Using $|X]=|Y]$ in the MHV vertex expansion,
there is no $z$-dependence in the MHV vertices, but each of the $k$ propagators falls off as $1/z$.
 Each MHV vertex diagram,
and hence the full amplitude, will then fall off as $1/z^k$.

\item
{\bf For even $n$, any amplitude
$\A^\text{N$^k$MHV}_n(z)$ goes as $O(z^{-k})$, or better, under an
 \emph{alternating shift}.}

\emph{Definition:} An alternating shift of an $n$-point amplitude with even $n$ is a square spinor shift
of the form~(\ref{generalsquare}) with $s=n/2$ lines shifted. Shifted and unshifted lines are chosen to alternate, \ie we  choose to shift either all even or all odd lines. As for common-index shifts, we impose the restriction~(\ref{commonindexci}) on the choice of shift parameters $c_i$.

\emph{Proof:}
As in the analysis of common-index shifts,  all ``end-vertices"  must contain shifted lines and~(\ref{commonindexci}) then implies that all propagators must shift.
In fact, each end-vertex must contain at least two consecutive external lines, and one of these lines shifts under the alternating shift.
It follows that each diagram falls off as $1/z^k$ for large $z$.

\item
{\bf Any amplitude
$\A^\text{N$^k$MHV}_n(z)$ goes as $O(z^{-k})$, or better, under an
 \emph{all-line shift}.}

\emph{Definition:}
All-line shifts were defined in~(\ref{commindexshift}) and are of the general form~(\ref{generalsquare}), with $m_i=i$ and $s=n$. Recall that the coefficients $c_i$ for all-line shifts satisfy~(\ref{subsetsumci}), \ie
\begin{equation}\label{alllineci}
    \sum_{i\in \alpha}c_i|i\>\neq0
\end{equation}
for all  ordered proper subsets $\alpha$ of the external states. Shift parameters $c_i$ satisfying~(\ref{alllineci}) are in fact sufficiently ``generic'' to ensure $1/z^k$ suppression of amplitudes, as we now show.

\emph{Proof:}
As in the analysis of alternating and common-index shifts, all ``end-vertices"  must contain shifted lines and thus all propagators shift. It follows from \reef{alllineci} that each diagram falls off as $1/z^k$ for large $z$. This verification of the $1/z^k$ falloff  directly from  the MHV vertex expansion is an important consistency check on our results in the previous sections.
\end{enumerate}

Any other square spinor shift that contains either all the alternating lines or a set of common-index lines also provides a $O(z^{-k})$ falloff for large $z$. Therefore any such shift gives rise to valid recursion relations for $k\ge 1$.

\setcounter{equation}{0}
\section{New form of the anti-NMHV generating function}\label{secanti}
The goal of this section is to obtain an improved version of the anti-NMHV generating
function presented in \cite{efk1} (see also \cite{sok2}).  This will lead to apparently new and curious
sum rules for the diagrams of both the NMHV and anti-NMHV generating functions.

\subsection{Anti-generating functions}
In the previous sections we have studied $\cn=4$ SYM amplitudes using the MHV vertex expansion. This expansion does not always yield the most convenient and efficient representation of an amplitude. For example, the MHV vertex expansion of an $n$-point N$^{k}$MHV amplitude
with $k=n-4$ generically contains many diagrams. But the amplitude is actually  anti-MHV so it can be
immediately expressed as the conjugate of an MHV amplitude.
More generally, since an N$^k$MHV amplitude is also anti-N$^{\bar k}$MHV with $\bar k=n-k-4$, an \emph{anti}-MHV vertex expansion is  more efficient when $\bar k<k$.

Generating functions for the anti-MHV vertex expansion of anti-N$^{\bar k}$MHV amplitudes can be readily obtained by \emph{conjugating} the generating functions~(\ref{FnNkMHV}) for N$^{k}$MHV amplitudes. The conjugate of a function $f(\<ij\>,[lm],\h_{ia})$ is simply $f([ji],\<ml\>, \bar{\h}_i^a)$,  defined with  reversed order of Grassmann monomials. However, the conjugated generating functions now depend on conjugate Grassmann variables $\bar{\h}_i^a$.
It is often more useful to re-express the generating function in terms of the original variables $\eta_{ia}$.
For example, we can then easily sum over the intermediate states of internal lines in unitarity cuts, even if the cut involves subamplitudes represented by both regular and conjugated generating functions.

To obtain the representation of the conjugate generating function in terms of $\eta_{ia}$'s, we use the Grassmann Fourier transform~\cite{sok2}
\be\lab{gft2}
\hat{f}([ji],\<ml\>,\eta_{ia}) ~\equiv~ \int \prod_{i,a} d\bar\h_i^a \,\exp\Bigl(\,\sum_{b,j}\h_{jb}\bar\h_j^b\Bigr)\,
f([ji],\<ml\>, \bar{\h}_i^a)\,.
\ee
The Fourier transform is equivalent to the following direct procedure~\cite{efk1} to obtain an anti-N$^{k}$MHV generating function from the corresponding  N$^{k}$MHV generating function:
\begin{enumerate}
\item Interchange all angle and square brackets: $\< ij \> \lra [ji]$.\vspace{-0.15cm}
\item Replace $\eta_{ia} \; \to \;\pa_i^a \;= \;\frac{\pa}{\pa \eta_{ia}}$.\vspace{-0.25cm}
\item Multiply the resulting expression by $\prod_{a=1}^4 \prod_{i=1}^n \eta_{ia}$ from the right.
\end{enumerate}
Applied   to the MHV generating function~(\ref{mhvgen}), this prescription gives the anti-MHV generating function~\cite{efk1}
\be  \lab{barFn}
\begin{split}
  \cf^{\overline{\rm MHV}}_n
  &=
  \frac{1}{2^4\,\cyc[1,\ldots,n]}
  \prod_{a=1}^4 \sum_{i,j\in{\rm ext} } \;[i j]\; \pa^a_i \pa^a_j ~
  \eta_{1a} \cdots \eta_{na} \\
  &=
  \frac{1}{(2\, (n-2)!)^4\,\cyc[1,\ldots,n]}
   \prod_{a=1}^4 \sum_{j_1,\ldots,j_n\in{\rm ext}}
   \eps^{j_1 j_2 \cdots j_n}\, [j_1 j_2]\, \eta_{j_3 a} \cdots \eta_{j_n a}\,,
\end{split}
\ee
where we write the conjugate of the cyclic product of angle brackets $\cyc(1,\ldots,n)$ as
\begin{equation}
    \cyc[1,\ldots,n]=(-1)^n\,\prod_{i=1}^n [i,i+1]\,.
\end{equation}

Similarly, we can apply the above prescription to obtain the  anti-NMHV generating function
$\cf^{\overline{\rm NMHV}}_{n}$ from its counterpart  ${\cf}^{\rm NMHV}_{n}$.
Recall that the latter is given by a sum over diagrams with subamplitudes $I_1$ and $I_2$, connected by an internal line of momentum $P_\alpha$, $\alpha=\{l_i,\ldots,l_f\}$.
The expansion \cite{efk1} takes the form (with $[P_\a| = \<X| P_\a$ and $\<X|$  a conjugate
reference spinor):
\begin{equation}\lab{barNFn0}
\begin{split}
  {\cf}^{\overline{\rm NMHV}}_{n}
  &=\sum_{{\rm diagrams}\,\alpha}
  \frac{1}{2^4\,\cyc[I_1]P_\a^2\,\cyc[I_2]} \;
  \prod_{a=1}^4 \sum_{i,j\in{\rm ext}}  \sum_{m \in \alpha}
  [i j] [ P_{\alpha}\,m] \; \pa_{i}^a\pa_{j}^a\pa_{m}^a\; \eta_{1a} \cdots \eta_{na} \\
  &=\frac{1}{\cyc[1,\ldots,,n]}\sum_{{\rm diagrams}\,\alpha}
  \frac{\overline{W}_{\alpha}}{(2\, (n-3)!)^4} \;
  \prod_{a=1}^4
  \sum_{m \in \alpha}\;
  \sum_{j_2,\dots,j_n\in{\rm ext} }
  [P_{\alpha} m] [j_2 j_3] \; \eps^{m j_2 \dots j_n}\;
  \eta_{j_4 a} \cdots \eta_{j_n a} \, .
\end{split}
\end{equation}
Here, the factor $\overline{W}_\alpha$ corresponding to a diagram with internal line $P_\alpha$ is given by~[\citen{ggk},\citen{efk1}]
\begin{equation}
    \overline{W}_\alpha=\frac{\cyc[1,\ldots,n]}{\cyc[I_1]P_\a^2\,\cyc[I_2]}
    =\frac{[l_i-1,l_i][l_f,l_f+1] }{[l_i-1,P_\alpha][l_i\,P_\alpha]\,P_\alpha^2\,[l_f\,P_\alpha][l_f+1,P_\alpha]}\,.
\end{equation}

\subsection{Extracting the overall Grassmann delta function}
From the explicit generating functions given in~(\ref{FnNkMHV}) and~\cite{Drummond:2008cr},  it is clear that any\footnote{
The only exception is the generating function for 3-point anti-MHV amplitudes.}
generating function can be written in the factorized form
\begin{equation}\label{extractd8}
    \cf_n=\delta^{(8)}\Bigl(\,\sum_{i\in{\rm ext}}|i\>\eta_{ia}\Bigr)\times\bigl[\ldots\bigr]\,.
\end{equation}
This form, with an explicit factor of $\delta^{(8)}$, is most useful in practical applications.
It is, for example, very convenient in the evaluation of unitarity cuts because the identity $\delta^{(8)}(X)\,\delta^{(8)}(Y)=\delta^{(8)}(X+Y)\,\delta^{(8)}(Y)$ can then be used to simplify calculations.
Furthermore, explicitly extracting the $\delta^{(8)}$ makes it manifest that $\cf_n$ is annihilated by the supercharge $\tilde Q_a$ defined in \reef{schg}. In fact, it follows from $X\delta^{(8)}(X)=0$ that
\begin{equation}\lab{susykill}
    \tilde Q_a\cf_n=-\Bigl( \sum_{i\in{\rm ext}}\<\epsilon i\>\eta_{ia}\Bigr) \delta^{(8)}\Bigl(\,\sum_{i\in{\rm ext}}|i\>\eta_{ib}\Bigr)\times\bigl[\ldots\bigr]=0\,.
\end{equation}
Therefore any generating function of $\cn=4$ SYM satisfies $\tilde Q_a\cf_n=0$.

 For the anti-MHV generating function~(\ref{barFn}),
 the desired form with
 the $\delta^{(8)}$ was recently presented in~\cite{efk1} (see also~\cite{sok2} for $n=4,5$):
\bea
  \lab{AbarFn}
  \cf^{\overline{\rm MHV}}_n
  \,=\,
  \frac{\d^{(8)} \bigl(\,\sum_{i\in{\rm ext}} |i\> \eta_{ia}\bigr)}
  {(2(n-4)!)^4\,\<12\>^4\, \cyc[1,\ldots,n]}
   ~   \prod_{a=1}^4 \sum_{j_3,\dots,j_n\in{\rm ext}}
   \eps^{1 2 j_3 \cdots j_n}\,
   [j_3 j_4]\, \eta_{j_5 a} \cdots \eta_{j_n a} \, .~~
\eea

To extract the $\delta^{(8)}$ in the anti-NMHV generating function~(\ref{barNFn0}), we start from its representation in the first line of~(\ref{barNFn0}), and use \reef{barFn} to obtain
\begin{equation}\label{NMHVfromMHV}
  {\cf}^{\overline{\rm NMHV}}_{n}
  =\sum_{{\rm diagrams}\,\alpha}
  \overline{W}_\alpha
  \Big(\prod_{a=1}^4 \sum_{m \in \alpha}  [ P_{\alpha}\,m]\,\pa_{m}^a\Big) \, {\cf}^{\overline{\rm MHV}}_{n}\,.
\end{equation}
We now use the form of $\cf^{\overline{\rm MHV}}_n$ in~(\ref{AbarFn}) and apply the product of $\pa_m^a$ derivatives.  These distribute themselves between the $\delta^{(8)}$-factor and the other factors of $\eta_{j_s a}$. We will prove in the next subsection that all terms in which the derivatives hit the $\d^{(8)}$ vanish. Thus we are left with the simple result
\begin{equation}
\boxed{\,
{\cf}^{\overline{\rm NMHV}}_{n} =\!\!\!
   \sum_{{\rm diagrams}\,\alpha}\!
    \frac{\overline{W}_{\alpha} \; \d^{(8)}\big(\sum_{i\in{\rm ext}} |i\> \eta_{ia}\big)}
    { (2 (n-5)!)^{4}\<12\>^{4}\cyc[1,\ldots,n] }
  \prod_{a=1}^4
  \sum_{m \in \alpha}
  \sum_{j_4,\dots,j_n \in {\rm ext}}\!\!\!\!\! \eps^{12mj_4\dots j_n}
   [m\, P_{\alpha}]  [j_4 j_5]\, \eta_{j_6 a} \cdots \eta_{j_n a}\,
   \,}
    \lab{antiNMHV}
\end{equation}
for the anti-NMHV generating function.
This form manifestly contains the desired Grassmann $\delta^{(8)}$ and is thus suitable for practical applications.
In fact, the generating function ${\cf}^{\overline{\rm NMHV}}_{n}$ in the form derived above will be used in\footnote{revised version, in preparation.}~\cite{efk1} to compute a generating function for 5-line unitarity cuts of 4-loop amplitudes.


\subsection{Proof of \reef{antiNMHV}}
Let us start with the expression \reef{NMHVfromMHV} and let the derivatives $\pa_m^a$ act. The result can be written (dropping non-vanishing overall  factors)
\bea
  \lab{GH1}
  {\cf}^{\overline{\rm NMHV}}_{n} ~\propto~\!\!
  \sum_{{\rm diagrams}\,\alpha}
    \overline{W}_{\alpha}
  \prod_{a=1}^4 \bigl(G^{\a}_a + P_\a^2 \,H_a\bigr) \, ,
\eea
where $P_\a^2\, H_a$ is the term\footnote{This is derived using
$ \half \sum_{m \in \a} [P_\a m]\,\pa^a_m \sum_{i,j \in {\rm ext}}\<ij\>\,\h_{ia}\h_{ja}= -P_\a^2 \sum_{i  \in {\rm ext}}\<X\,i\>\,\h_{ia}$.}
 that arises when  the derivative $\pa^a_m$ hits the $\delta^{(8)}$, and $G^\alpha_a$ the term when it hits the other factor of \reef{AbarFn}. Specifically we have
\bea
  G^\alpha_a &=& (n-4) \Big( \sum_{i,j\in{\rm ext}} \<ij\>\, \eta_{ia} \eta_{ja}\Big)
    \sum_{m \in \alpha}\,\,
  \sum_{j_4,\dots,j_n \in {\rm ext}} \!\!\!\eps^{12mj_4\dots j_n}
   [P_{\alpha} \, m]  [j_4 j_5]\, \eta_{j_6 a} \cdots \eta_{j_n a}\, ,\\[2mm]
  H_a &=&  {}-2 \, \Big[\sum_{i\in{\rm ext}}\<X\,i\> \h_{ia}\!\!\!\sum_{j_3,\ldots, j_n\in{\rm ext}}\!\!\! \e^{12j_3j_4\ldots j_n}[j_3\,j_4]  \h_{j_5a}\ldots\h_{j_na} \Big]\,.
\eea
Note that $H_a$ is manifestly independent of the diagram $\alpha$.

The action of the supercharge $\tQ_a$ on
 $G_a^\alpha$ and $H_a$ will be important in the following.
We  immediately see that $G_a^\alpha$ is annihilated by $\<Y\tilde{Q}_a\>$
for all choices of the angle spinor $\<Y|$:
\begin{equation} \lab{gkill}
    \<Y \tilde{Q}_a\>G^\alpha_a=0\,.
\end{equation}
Here and in the following, repeated indices $a$ are \emph{not} summed over.
Using the relation
\be \lab{simp}
\<Y \tilde{Q}_a\>\sum_{i\in{\rm ext}} \<X i\>\eta_{ia}~=~\frac12 \,\<Y\,X\>\!\!\sum_{i,j\in {\rm ext}}\<ij\>\,\h_{ia}\h_{ja}\,,
\ee
which follows from the Schouten identity, we compute
\begin{equation}
  \<Y \tilde{Q}_a\>H_a= \<XY\>\, \Big[\sum_{i,j\in{\rm ext}}\<i\,j\> \h_{ia}\h_{ja}\!\!\!\sum_{j_3,\ldots, j_n\in{\rm ext}}\!\!\! \e^{12j_3j_4\ldots j_n}[j_3\,j_4]  \h_{j_5a}\ldots\h_{j_na} \Big]\,.
\end{equation}
If $\<XY\>\neq0$, the right hand side is manifestly non-vanishing. This can seen by eliminating its dependence on the Grassmann variables $\eta_{ia}$ through differentiation:
\begin{equation}
    \pa_{1}^a\pa_{2}^a\,\pa_{5}^a\cdots \pa_{n}^a\<Y \tilde{Q}_a\>H_a=4(n-4)!\,\<XY\>\<12\>[34]\neq0\,.
\end{equation}

Let us now expand out the product in \reef{GH1},
\begin{equation}  \lab{GH2}
\begin{split}
  {\cf}^{\overline{\rm NMHV}}_{n} &\propto\!\!
  \sum_{{\rm diagrams}\,\alpha}
    \overline{W}_{\alpha}
  \prod_{a=1}^4 (G^{\a}_a + P_\alpha^2\,H_a)\\
   &=\!\!\!
   \sum_{{\rm diagrams}\,\alpha}
    \overline{W}_{\alpha}
    \bigg(
   \prod_{a=1}^4 G^\alpha_a +  \bigl(P_\alpha^2\bigr)^4\prod_{a=1}^4 H_a
   + \bigl(P_\alpha^2\bigr)^3\bigl( G^\alpha_1 H_2 H_3 H_4 + \rom{perms}\bigr) \\[1mm]
   &~~~~~~~~~~~~~~~~~~~~~+ \bigl(P_\alpha^2\bigr)^2\bigl( G^\alpha_1 G^\alpha_2 H_3 H_4 + \rom{perms}\bigr)
   + P_\alpha^2\,\bigl( G^\alpha_1 G^\alpha_2 G^\alpha_3 H_4 + \rom{perms}\bigr)
   \bigg)
    \, .
\end{split}
\end{equation}
The first term $\prod_{a=1}^4 G^\alpha_a$ has an explicit factor of
$\delta^{(8)}$ and it is (once overall factors are restored) the desired generating function \reef{antiNMHV}. We will argue that each of the other terms vanishes.

By virtue of \reef{extractd8} and \reef{susykill},  the generating function~(\ref{GH2}) must be annihilated by $\<Y\tilde{Q}_a\>$ for all $a$.   Then \reef{gkill} implies that
\begin{equation}\label{H4zero}
      \biggl[\,\prod_{a=1}^4\<Y\tilde{Q}_a\>\biggr]{\cf}^{\overline{\rm NMHV}}_{n}\,=\,0
   \qquad\Longrightarrow\qquad
    \Biggl[\,\sum_{{\rm diagrams}\,\alpha} \overline{W}_{\alpha}\bigl(P_\alpha^2\bigr)^4\Biggr]\,\prod_{a=1}^4 \<Y \tilde{Q}_a\>H_a  ~=~0\, .
\end{equation}
As each $\<Y \tilde{Q}_a\>H_a$ is manifestly non-vanishing, their products must also be non-vanishing.\footnote{Each factor is a polynomial of \emph{different} Grassmann variables. Therefore the product is non-vanishing as long as all factors are non-vanishing.}
We conclude that~(\ref{H4zero}) implies the sum rule
\begin{equation}\label{sumrule}
    \sum_{{\rm diagrams}\,\alpha}
  \overline{W}_\alpha
   \bigl(P^2_{\alpha}\bigr)^4=0\,.
\end{equation}
It follows from~(\ref{sumrule}) that the term containing $\prod_{a=1}^4 H_a$ in the anti-NMHV generating function \reef{GH2} vanishes.

We turn to the terms with three factors of $H_a$. For definiteness, consider the term with $G^\alpha_1 H_2 H_3 H_4$. This time we act with three SUSY transformations and use the vanishing of the term $\prod_{a=1}^4 H_a$ established above to obtain
\begin{equation}\label{H3zero}
      \biggl[\,\prod_{a=2}^4\<Y\tilde{Q}_a\>\biggr]{\cf}^{\overline{\rm NMHV}}_{n} \,=\,0
      \qquad\Longrightarrow\qquad
    \Biggl[\,\sum_{{\rm diagrams}\,\alpha} \overline{W}_{\alpha}\bigl(P_\alpha^2\bigr)^3G_1^\alpha\biggr]\,\prod_{a=2}^4 \<Y \tilde{Q}_a\>H_a~=~0\,.
\end{equation}
SUSY invariance thus implies the sum rule
\begin{equation}\label{sumrule2}
    \sum_{{\rm diagrams}\,\alpha} \overline{W}_{\alpha}\bigl(P_\alpha^2\bigr)^3G_a^\alpha=0\,.
\end{equation}
 Note that multiple relations are implied by~(\ref{sumrule2}), namely one for each distinct Grassmann monomial appearing on the left hand side.

The relation~(\ref{sumrule2}) immediately implies the vanishing of the terms involving $G^\alpha_1 H_2 H_3 H_4$ and its permutations in ${\cf}^{\overline{\rm NMHV}}_{n}$. With $H^4$ and $GHHH$ already argued to be absent, we can proceed and act with two SUSY transformations to isolate terms of the form $GGHH$. SUSY invariance gives another sum rule
\begin{equation}\label{sumrule3}
    \sum_{{\rm diagrams}\,\alpha} \overline{W}_{\alpha}\bigl(P_\alpha^2\bigr)^2G_a^\alpha G_b^\alpha=0\,,
\end{equation}
and this sum rule in turn implies that terms of the form $GGHH$ must vanish. Finally, the vanishing of terms of the form $GGGH$ follows from the sum rule
\begin{equation}\label{sumrule4}
    \sum_{{\rm diagrams}\,\alpha} \overline{W}_{\alpha} P_\alpha^2\, G_a^\alpha G_b^\alpha G_c^\alpha=0\,.
\end{equation}
This completes the derivation of the anti-NMHV generating function~(\ref{antiNMHV}).
In the next subsection we discuss the four sum rules which arose in the course of the proof.

\subsection{Sum rules}\label{ssecsumrule}
 In the previous section we have derived the sum rules~(\ref{sumrule}),~(\ref{sumrule2}),~(\ref{sumrule3}), and~(\ref{sumrule4}). We will now argue that, except for~(\ref{sumrule}), all of these relations can be independently derived
 from Cauchy's theorem as sum rules associated with certain square spinor shifts.

Let us start with the sum rule~(\ref{sumrule2}). Substituting $G_a^\alpha$, we find
\begin{equation}\label{expsumrule2}
    \sum_{{\rm diagrams}\,\alpha} \overline{W}_{\alpha}\bigl(P_\alpha^2\bigr)^3
\Big(\sum_{i,j\in{\rm ext}} \<ij\>\, \eta_{ia} \eta_{ja}\Big) \sum_{m \in \alpha}\,\,
  \sum_{j_4,\dots,j_n \in {\rm ext}} \!\!\!\eps^{12mj_4\dots j_n}
   [P_{\alpha}\, m]  [j_4 j_5]\, \eta_{j_6 a} \cdots \eta_{j_n a} =0\,.
\end{equation}

Now consider a square bracket shift of the 3 negative helicity lines $m_{1,2,3}$ of any $n$-point gluon NMHV amplitude $\A^{\rm NMHV}_n$. The shift can be written in the form
\begin{equation}
    |\hat m_1]= |m_1]+ z \<m_2m_3\>|X]\,,\quad
    |\hat m_2]= |m_2]+ z \<m_3m_1\>|X]\,,\quad
    |\hat m_3]= |m_3]+ z \<m_1m_2\>|X]\,,
\end{equation}
which, by the Schouten identity, satisfies the momentum conservation condition~(\ref{sumcisquare}).
Let $\A^{\rm NMHV}_n(z)$ denote the shifted amplitude.
As $z \to \infty$, a typical gluon amplitude falls off as $1/z^{4}$ and we obtain the contour integral identity
\begin{equation}\label{gensumrule}
    \frac{1}{2\pi i}\oint \frac{dz}{z}\, \A^{\rm NMHV}_n\!(z)\,z^q =0\,\qquad\text{ for }\quad q\leq3\,.
\end{equation}
When $q=0$, this gives the usual MHV vertex expansion of $\A^{\rm NMHV}_n$. When $q=1,2,3$\,,
 there is no pole at $z=0$
and instead we get the set of sum rules
\bea
 0~=\sum_\alpha{}^\prime~ W_\a \< m_2 m_3 \>^4 \< m_1 P_\a \>^4 \, z_{\a}^q\, .
\eea
The designation $\sum'$ means that the sum includes all poles of the shifted amplitude. Thus the
sum is over diagrams $\a$ where the negative helicity lines $m_i$ are distributed with $m_1$ on one MHV vertex and $m_2$ and $m_3$ on the other MHV vertex. The value of $z$ at the pole with channel momentum $P_\a$ is $z_\a = P_\a^2 / (\< m_2 m_3 \> \< m_1 P_\a \>)$, so the sum rules can also be written
\bea\label{SR0}
   0~=\sum_\alpha{}^\prime~ W_\a \< m_2 m_3 \>^{4-q} \< m_1 P_\a \>^{4-q} \, (P_{\a}^2)^q\, .
\eea
The conjugate versions of these sum rules are
\bea
  \lab{SR}
  0~=\sum_\alpha{}^\prime~ \overline{W}_\a [ m_2 m_3 ]^{4-q} [ m_1 P_\a ]^{4-q} \, (P_{\a}^2)^q\, .
\eea
Let us now compare this to~(\ref{expsumrule2}).
Each Grassmann monomial in~(\ref{expsumrule2}) can be seen to be equivalent to the sum rule~(\ref{SR}) with $q=3$ for one particular choice of NMHV gluon amplitude.
For example, we can project onto the monomial $\eta_{1a}\eta_{2a}\eta_{6a}\cdots\eta_{na}$ by acting with Grassmann derivatives $\pa_1^a\pa_2^a\pa_6^a\cdots \pa_n^a$ on~(\ref{expsumrule2}). The resulting relation is of type~(\ref{SR}), derived as explained above by conjugating the sum rule associated with the NMHV gluon amplitude $\A^{\rm NMHV}_n(\ldots,3^-,4^-,5^-,\ldots)$\,.

The remaining relations~(\ref{sumrule3}) and~(\ref{sumrule4}) similarly arise from sum rules of the form~(\ref{gensumrule}), applied to NMHV amplitudes with more generic external states of the $\cn=4$ theory. In fact,~(\ref{sumrule3}) arises from $q=2$ sum rules that are associated with shifts of three negative helicity gluinos that carry the same three $SU(4)$ indices. The fourth $SU(4)$ index is distributed arbitrarily.\footnote{If the fourth index happens to also sit on a shifted line, the corresponding particle is of course a gluon instead of a gluino.}  Such a shift generically gives $1/z^3$ suppression, thus validating the sum rule. Finally,~(\ref{sumrule4}) arises from $q=1$ sum rules associated with shifts of three scalars which two common $SU(4)$ indices, while there is no restriction on the remaining indices.
This time we only have $1/z^2$ suppression under the shift, which still justifies the $q=1$ sum rule.

The sum rule~(\ref{sumrule})
 is intriguingly simple.\footnote{As a check on the derivation by SUSY invariance in section
 8.3, we have verified the sum rule numerically up to $n=11$ external legs.}
Its conjugate version is given by
\begin{equation}\label{conjsumrule}
\boxed{\phantom{\biggl(}
    \sum_{{\rm diagrams}\,\alpha}
   W_\alpha
   \bigl(P^2_{\alpha}\bigr)^4=0\,
   \quad\text{with}\quad
   W_\alpha=\frac{\cyc(1,\ldots,n)}{\cyc(I_1)\,P_\alpha^2\,\cyc(I_2)}\,.
   ~}
\end{equation}
Formally, this sum rule takes the form~(\ref{SR0}) with $q=4$, but notice that the sum over diagrams $\alpha$ in~(\ref{conjsumrule}) goes over \emph{all possible diagrams}, while~(\ref{SR0}) is limited to the diagrams in the expansion from the shift under consideration.
Thus, in general, the sum rule (\ref{conjsumrule}) cannot be derived directly from \reef{SR0}.
The cases  $n=5,6$ external lines are an exception. For $n=6$, for example,~(\ref{conjsumrule}) can be derived from a sum rule of the form~(\ref{SR0}) for the alternating helicity amplitude $A_6^{\rm NMHV}(1^-,2^+,3^-,4^+,5^-,6^+)$.
In this case, all possible diagrams contribute to the recursion relation following from the square spinor shift of the 3 negative helicity lines.
This enables it to reproduce~(\ref{conjsumrule}) for $n=6$. Furthermore the shift provides $1/z^5$ suppression, validating its $q=4$ sum rule.

For $n>6$, the diagrams of three-line shifts never include all possible channels $\alpha$. Furthermore, the $q=4$ sum rules are generically not valid, because any NMHV gluon amplitude goes as $1/z^4$ under a shift of negative helicity gluons, and thus $\A^{\rm NMHV}_{n}(z)\, z^4$ has a pole at infinity. Curiously, we now have a simple way of computing the residue of this pole at infinity! As~(\ref{conjsumrule}) must hold, the residue at $z=\infty$ is simply the sum over all other remaining diagrams which are \emph{not} in the expansion (denoted by $\sum''$):
\begin{equation}\label{}
    {}-\frac{1}{2\pi i}\oint \frac{dz}{z}\, \A^{\rm NMHV}_{n}\!(z)\,z^4
    ~=~
    \sum_\alpha{}^{\prime\prime}~
   W_\alpha
   \bigl(P^2_{\alpha}\bigr)^4\,.
\end{equation}
Since the sum rule~(\ref{conjsumrule}) is so simple and appears to be related to pure-gluon amplitudes,
it would be interesting to find a more conventional derivation rather than the route
we took using
SUSY invariance of generating functions.



\setcounter{equation}{0}
\section{Discussion}
\lab{s:disc}

In this paper, we have proven that the MHV vertex expansion is valid for all amplitudes of the $\cn=4$ SYM theory, and we have presented compact expressions for generating functions which efficiently encode the dependence on external states for all N$^k$MHV amplitudes of the theory.

Our proof of the MHV vertex expansion used recursion relations associated with ``all-line shifts", which involved complex deformations of the square spinors of the momenta of all external lines. We established the validity of these recursion relations  by showing that N$^k$MHV amplitudes vanish at large $z$ under all-line shifts for all $k\geq1$. We showed that the all-line shift recursion relations iterate to give precisely the MHV vertex expansion, and this allowed us to derive the compact expressions (\ref{FnNkMHV}) for the generating functions for N$^k$MHV amplitudes.

In unitarity cuts of loop amplitudes, one must sum over all possible intermediates states that can run in the loops. The generating functions  developed here provide a systematic way to efficiently perform such intermediate state sums in $\cn=4$ SYM when the cuts reduce the loop amplitude to products of tree amplitudes.
In fact, the generating functions can be used in the computation of the state sum even before particles  are assigned to the external lines of the loop amplitude. The result is a generating
function for the particular unitarity cut of the amplitude. Several non-trivial examples of intermediate state sums for cut 1-, 2-, 3-, and 4-loop amplitudes were carried out in\cite{efk1} using MHV and NMHV generating functions and their conjugate anti-generating functions.
As higher loop-level requires the inclusion of N$^k$MHV subamplitudes of higher $k$, the methods of~\cite{efk1} can now be combined with
the generating functions~(\ref{FnNkMHV}) to evaluate state sums for  unitarity cuts of $L\geq 4$ loop amplitudes.

The KLT relations express tree amplitudes of $\cn=8$ supergravity as sums of products of
two $\cn=4$ SYM tree amplitudes. Thus results on $\cn=4$ trees can be applied in studies of
supergravity. In particular intermediate state sums for unitarty cuts in supergravity
can be performed as the product of unitarity sums in SYM \cite{Bern:2008pv}.
The perturbative UV behavior  of $\cn=8$ supergravity has recently been under intense investigation~\cite{Green:2006gt,Green:2006yu,Bern:2006kd,Bern:2007hh,unexp,Bern:2008pv,Naculich:2008ew}, and surprising structures at loop-level, such as the absence of triangle and bubble graphs~[\citen{Bern:2005bb,BjerrumBohr:2006yw,BjerrumBohr:2008ji,BjerrumBohr:2008vc},\citen{nima2}], seem to play an important role in the cancellation of divergences.

Although connected through the KLT relations, gravity and gauge theory amplitudes may behave differently under shifts.
Notably, even if all gauge theory  amplitudes in the KLT formula vanish at large $z$ under a given shift of the external momenta, this shift does not necessarily imply a valid recursion relation for the supergravity amplitude. The reason is that in the KLT relations for $n$-point supergravity amplitudes each product of gauge amplitudes is multiplied by a product of $n-3$ Mandelstam variables, which can grow as fast as $z^{n-3}$ under the shift. The shifted supergravity amplitude may then not vanish at large $z$. To be specific, consider an $n$-point graviton NMHV amplitude. KLT expresses it in terms of products of gluon NMHV amplitudes. Under a shift of the three negative helicity gravitons, the gluon amplitudes needed in KLT generically have $1/z^4$ falloffs, giving a large $z$ suppression of $(1/z^4)^2$. But the $n-3$ Mandalstam variables also shift to give $z^{n-3}$, and the naive large $z$ behavior is thus $z^{n-11}$; numerical tests (for $n=5,\dots,11$) show that one
power of $z$ cancels when all terms in the KLT relation are summed, giving the leading $z^{n-12}$ behavior first reported in \cite{BEF}.
The failure of the large $z$ suppression for $n \ge 12$ means that there is an extra term from the pole at infinity contributing to the recursion relation, which for this shift is the MHV vertex expansion \cite{greatdane,BEF}.
Similar problems arise for lower-point amplitudes when the external gravitons are replaced by other states of the $\cn = 8$ theory.
These examples illustrate that the validity of the MHV vertex expansion is a non-trivial statement. A proof for $\cn = 4$ is thus a necessary prerequisite for using confidently the MHV vertex approach in applications, such as the intermediate state sums.

We showed in sections \ref{secvalidity} and \ref{secsquare} that
three types of square spinor shifts provide (at least) $1/z^k$ large $z$ suppression for $\cn =4$  N$^k$MHV amplitudes; these shifts are the
common-index shift, the alternating shift, and the all-line shift (see section \ref{secsquare} for their precise definitions). For sufficiently large $k$, these shifts thus give enough suppression in the gauge theory amplitudes to counter the large $z$ divergences of the Mandelstam variables in the KLT relations, and they then
imply valid recursion relations for supergravity amplitudes.
To be concrete, consider an $n$-point N$^k$MHV supergravity amplitude $\mathcal{M}_n$ under an all-line shift. We need $n\ge k+4$ in order for the amplitude to be non-vanishing. KLT relates the amplitude to squares of $n$-point N$^k$MHV gauge theory amplitudes, which fall off as $(1/z^k)^2$, or better, under the shift. The Mandelstam factors give at worst $z^{n-3}$. So for $k+4 \le n<2k +3$, the supergravity amplitude vanishes\footnote{The upper bound on $n$ may be too strong, since for particular amplitudes, the falloff of the gauge theory amplitudes could be better and the because there could be cancellations between terms in the KLT relations.} for large $z$, and the resulting recursion relations provide a valid representation of $\mathcal{M}_n$.
It would be interesting to see whether these recursion relations offer useful and efficient ways to compute supergravity amplitudes.

In section \ref{ssecsumrule}, we derived sum rules which are closely related to the fast large $z$ falloff of amplitudes under common-index shifts. It seems to be the general rule of thumb that the more diagrams in a recursion relation, the faster the falloff under the associated shift, and hence the more relations between the diagrams.\footnote{We thank F.~Cachazo for discussions of this point.} 
Also, in section \ref{ssecsumrule} we derived a new sum rule which included a sum over all MHV vertex diagrams at the NMHV level. It would be interesting to see if this sum rule can be derived from shifts, and whether there exists a simple generalization at the N$^k$MHV level.

In~\cite{nima2}  and  \cite{Brandhuber:2008pf} it was proposed to apply shifts directly to generating functions and supplement the BCFW 2-line shift by a shift of Grassmann variables.
The supercharge $|\tQ_a \> = \sum_i |i\>\h_{ia}$ is invariant under the combined shift, which one may call a \emph{supershift}.
This leads directly to recursion relations for the generating functions.
One can define supershifts which supplement the \emph{square spinor} shifts we have used, and it could be worthwhile to consider this idea more carefully.

The 2-line supershifts were used to prove dual superconformal symmetry
\cite{Drummond:2006rz,Alday:2007hr,dks,Drummond:2007cf,Drummond:2007au,dhks,Alday:2007he,Brandhuber:2008pf,Beisert:2008iq,Berkovits:2008ic,McGreevy:2008zy,sok2} of tree level amplitudes \cite{Brandhuber:2008pf}.  In fact, the supershift recursion relations can be solved,  giving a construction of  generating functions for all N$^k$MHV amplitudes of $\cn=4$ SYM \cite{Drummond:2008cr}.  The generating functions are expressed in terms of superconformal invariants, with
new invariants required at each level $k$.  At the NMHV level the generating function of
\cite{Drummond:2008cr} corresponds to a sum of  $(n-3)(n-4)/2$ diagrams while
$n(n-3)/2$  MHV vertex diagrams needed  at the same level. Both types of generating functions can be used in the evaluation of amplitudes or unitarity cuts. It would be interesting to compare the two approaches in practical applications.

There has been recent progress in \emph{off-shell} approaches to MHV expansions \cite{Mansfield:2005yd,Vaman:2005dt,Feng:2006yy,Boels:2006ir,Boels:2007qn,Ananth:2007zy,Mason:2008jy}. The idea is to construct Lagrangians in which every interaction vertex is MHV. The on-shell and off-shell MHV methods lead to very similar diagrammatic expansions.
A more detailed comparison of the results would be interesting.

\section*{Acknowledgements}

We are grateful to Nima Arkani-Hamed, Zvi Bern, Emil Bjerrum-Bohr, Freddy Cachazo, Lance Dixon, Stephen Naculich and Pierre Vanhove for valuable discussions. MK would like to thank the Institute for Advanced Study for hospitality during the final stages of this work.
HE is supported by NSF grant PHY-0503584.
DZF is supported by NSF grant PHY-0600465.
DZ and MK are supported by the US  Department of Energy through cooperative research agreement DE-FG0205ER41360.

\appendix
\setcounter{equation}{0}
\section{Proof of~(\ref{deltaID})}
\label{appdeltaID}
In this appendix we prove the identity~(\ref{deltaID}), namely
\begin{equation}  \lab{bigide}
    \prod_{B=1}^{k+1}\delta^{(8)}\Big(I_B\Bigr)
    =\delta^{(8)}\biggl(\sum_{i\in{\rm ext}}|i\>\eta_{ia}\biggr)
    \prod_{A=1}^{k}\delta^{(8)}\biggl(\,\sum_{i\in\alpha_A}|i\>\eta_{ia}-|P_{\alpha_A}\>\eta_{\alpha_{\!A}a}\biggr)\,,
\end{equation}
where $\{\alpha_1,\ldots,\alpha_k\}$ and $I_1,\ldots,I_{k+1}$ characterize, respectively, the internal lines and subamplitudes of an MHV vertex diagram.
We will refer to the subamplitudes as vertices.

The proof is facilitated if we order the $k+1$ vertices $I_B$ and choose the $k$ channels $\a_A$  for each diagram in the following way:
\begin{enumerate}
  \item Any diagram contains
  $s\geq2$
  ``end-vertices'', \ie vertices which only connect to one internal line. Start with these and label them
   $ I_1, I_2,\dots,I_s$.   Label the adjacent internal lines with channel labels $\a_1,
\a_2, \dots, \a_s$. The direction of  channel momenta  is chosen so that
$P_{\a_A} = \sum_{{\rm ext} i\, \in I_A} p_i$.
  \item
  If not all lines are labeled in step 1, we proceed as follows.
  Consider all vertices which each
  contain a) an arbitrary number of external lines,
  b) one or more
of the internal lines labeled in step 1, and c) exactly  one internal line not yet labeled. Suppose that there are $t$ such vertices.
 It is easy to see that $t\geq2$.
 Label them $I_{s+1}, I_{s+2},\dots,I_{s+t}$.
Label the $t$  new internal lines respectively as  $ \a_{s+1}, \a_{s+2},\dots,\a_{s+t}$.  Again choose the direction of channel momenta so that
 the external states of $I_A$ are contained in $\alpha_A$.
  \item Continue this procedure including
  the two or more
  new vertices which contain external lines, previously labeled internal lines,
  and one unlabeled line
  at each stage until all
  internal lines are labeled.
 \item Finally, we label the last remaining unlabeled vertex by $I_{k+1}$.
\end{enumerate}

In order to rewrite the $\d$-functions on the LHS of \reef{bigide}, we use repeatedly the Grassmann $\d$-function identity
\bea
\lab{dID}
\delta^{(8)}(X)\,\delta^{(8)}(Y)=\delta^{(8)}(X)\,\delta^{(8)}(X+Y) \, .
\eea
Start with
 $\d^{(8)}(I_{k+1})$. Using all the $k$ other $\d$-functions in the product on the LHS of \reef{bigide}, we can replace
$\d^{(8)}(I_{k+1})$ by
 $\d^{(8)}(\sum_{A=1}^{k+1} I_A ) = \bigl(\sum_{i\in{\rm ext}}|i\>\eta_{ia}\bigr)$. This is the overall $\d$-factor on the RHS of  \reef{bigide}. We have used that the internal line contributions $|\pm P_{\a_A}\> \eta_{\alpha_A a}$ cancel pairwise and that each external leg sits on precisely one vertex.

Note that the $\d$-functions for the end-vertices $I_{1}, I_{2},\dots,I_{s}$ of step 1 are already in the form that appears on the RHS of \reef{bigide}. We now argue that the remaining $\d$-functions can also be converted to this form.
Consider the vertices $I_{s+1}, I_{s+2},\dots,I_{s+t}$ of step 2 above.
Use \reef{dID} to eliminate their dependence on the internal lines that connect each of them to the end-vertices of step 1. As a concrete example, we have\footnote{
The Kronecker delta $\delta_{\alpha_{\!A}\in I_B}$ is 1 if the internal line $P_{\alpha_A}$ is contained $I_B$, and vanishes otherwise.}
\bea
  \d^{(8)}(I_{s+1}) ~\to~
  \d^{(8)}\Bigl(I_{s+1} + \sum_{A=1}^s \delta_{\a_A \in I_{s+1} }\; I_{A}\Bigr)
   = \d^{(8)}\Big(\,\sum_{i\in\alpha_{s+1}}|i\>\eta_{ia}-|P_{\alpha_{s+1}}\>\eta_{\alpha_{s+1}a}\Big) \, ,
   \lab{Isp1}
\eea
where $P_{\alpha_{s+1}}$ denotes the  single momentum line of $I_{s+1}$ that does not connect to an end-vertex. The terms $|P_\a\> \eta_{\a}$ of the internal lines that connect $I_{s+1}$ to end-vertices are canceled. Since momentum is conserved\footnote{In the $\d$-functions on the LHS of \reef{bigide}, the momenta associated with the angle spinors add up to zero. It is clear that the identity \reef{dID} preserves this property. It must therefore also hold on the RHS of \reef{Isp1}.} on each of the vertices, the sum over external legs on the RHS of \reef{Isp1} are precisely those that add up to the remaining internal line $P_{\alpha_{s+1}}$.
This way all $t$ $\d$-functions for the vertices in step 2 are rewritten in the form needed for the RHS of \reef{bigide}.

When manipulating the vertices labeled in step 3, one can at each iteration use the previously rewritten $\d$-functions. For instance, for $\d^{(8)}(I_{s+t+1})$ we can use the first $s+t$ $\d$-functions to eliminate the dependence of internal momenta connecting $I_{s+t+1}$ to the $I_{A}$'s with $A=1,\dots,s+t$. It is clear that the end-result is the formula \reef{bigide}.

The labeling of channels $\alpha_A$ and vertices $I_B$ described by the steps 1\,--\,4 above was convenient to carry out the proof. It is clear, however, that formula \reef{bigide} holds independent of this particular prescription because it is invariant under a relabeling of channels and vertices.

\section{Kinematics of the shifts}
\lab{app:shifts}

In our proof, we will perform a square spinor shift
of all external lines on the BCFW representation of the amplitude, so we need details of the kinematics of both the BCFW shift and the subsequent all-line shift.

\subsubsection*{BCFW shift (primary)}
The $[1,\ell\>$-shift is defined as
\bea
  |\tilde{1}] = |1] + z |\ell] \, ,~~~~ |\tilde{1}\> = |1\> \, , ~~~~~~~~~~~
  |\tilde{\ell}] = |\ell] \, ,~~~~|\tilde{\ell}\> = |\ell\> - z |1\> \,.
\eea
Consider a diagram of the $[1,\ell\>$-expansion with internal momentum $\tilde{P}_{I}$, defined as the sum of external momenta on the left subamplitude $I$.
The condition that the internal momentum is on-shell fixes the value of $z$ at the pole to be
$z_{I}=P_{I}^2/\<1|P_I |\ell]$. The shifted
spinors
at the pole are
\bea
  |\tilde{1}] ~=~ |1] + \frac{P_{I}^2}{\<1|P_I |\ell]} |\ell] \, ,
  \hspace{1cm}
  |\tilde{\ell}\> ~=~ |\ell\> -  \frac{P_{I}^2}{\<1|P_I |\ell]} |1\> \, .
\eea
The internal momentum
$(\tilde{P}_{I})^{\da\b}= P_{I}|\ell]\,\<1|P_{I}/\<1|P_I |\ell]$ is on-shell (null), and it is convenient to define spinors associated with $\tilde{P}_I$ as
\bea
  |\tilde{P}_I\> ~=~  \frac{\< 1\ell \>}{\<1|P_I |\ell]}\; P_I |\ell]\, ,
   \hspace{1cm}
  [ \tilde{P}_I | ~=~ \frac{\<1|P_I}{\< 1\ell \>} \, .
\eea

\subsubsection*{All-line shift (secondary)}

We now consider the kinematics of a shift of the square spinors of all external lines, an all-line
shift. This shift is performed on the individual BCFW diagrams, which contain the shifted lines $\tilde{1}$, $\tilde{\ell}$ and $\tilde{P}_I$. The channel momentum $\tilde{P}_I$ is defined as the sum of external momenta on the left subamplitude $I$, which we define to be the vertex which contains the the BCFW-shifted line $\tilde{1}$, see figure \ref{figBCFW}. The right subamplitude, $J$, then contains $\tilde{\ell}$ as one of its external states.

Recall that an
all-line shift
acts as
\bea
  \lab{allshift}
  |\,\hat{i}\,] = |\,i\,] + z \, c_i \, |X] \, ,\hspace{8mm}
  \text{with} \hspace{8mm}
  \sum_{i \in {\rm  ext}} c_i\, \<\,i\,|  = 0 \, ,
\eea
on all external lines $i=1,\dots,n$.
The channel momentum $P_I$ shifts as
\begin{equation}\label{shiftedPI2}
    \hat P_{I}^2=P_{I}^2-z  \sum_{i \in I} c_i\, \<i|P_I|X]\,,
\end{equation}
where it is understood that the sum is over the external legs of $I$.
We will write
\begin{equation}
  \hat{P}_I |\hat{\ell}] ~=~ P_I |\ell] + z\, | \mathring{P}_I \>\
\end{equation}
with
\begin{equation}
  | \mathring{P}_I \> ~\equiv~
  c_\ell \,  P_I |X]
  - \sum_{i \in I} c_i \,| i \>  [\ell X]
  ~=~
  -c_\ell \, (P_J - \ell) |X]
  + \sum_{j \in J \backslash \{\ell\}} c_j \,| j \>  [\ell X] \, ,
\end{equation}
where $P_J = - P_I$ and we have used the momentum conservation condition in \reef{allshift}.

As $z \to \infty$, the leading behavior of $|\hat{\tilde{P}}_{I}\>$ and $|\hat{\tilde{P}}_{I}]$ is given by
\begin{equation}
\begin{split}\label{shiftedPI}
  |\hat{\tilde{P}}_{I}\>
  ~&=~  \frac{\< 1\ell \>}{\<1|\hat{P}_I|\hat{\ell}]}\; \hat P_{I}|\hat{\ell}]
  ~= ~\frac{\< 1\ell \>}{\<1 \mathring{P}_I \>}\, | \mathring{P}_I \>  +O(z^{-1})\\
  | \hat{\tilde{P}}_I ] ~&=~ \frac{\<1|\hat P_{I}}{\< 1\ell \>}
  ~= ~z\,\tilde{c}_P\,|X]+O(1)
  \qquad\text{with}\quad\tilde{c}_P = \sum_{i\in I} c_i\, \frac{\< 1 i \>}{\< 1 \ell \>}
  \,.
\end{split}
\end{equation}

The all-line shift also affects lines $\hat{1}$ and $\hat{\ell}$; we find
\begin{equation}
\begin{split}
  |\hat{\tilde{1}}]
  ~&=~ |\hat{1}] + \frac{\hat{P}_I^2}{\<1|\hat{P}_I|\hat{\ell}]} \;|\hat{\ell}]
  \,~=~ z\,\tilde{c}_1\,
  |X]
  +O(1)
  \qquad\text{with}\quad
  \tilde{c}_1 = c_1 - c_\ell\, \frac{\sum_{i \in I} c_i \< i | P_I |X]}{\<1 \mathring{P}_I \>}
  \,,\\
  |\hat{\tilde{\ell}}\>
  ~&=~ |\ell\> -  \frac{\hat{P}_I^2}{\<1|\hat{P}_I|\hat{\ell}]}\; |1\>
 ~= ~ |\ell\> +  \frac{\sum_{i\in I} c_i \<i|P_I |X]}{\<1\mathring{P}_I\>}\; |1\> +O(z^{-1})\, .
\end{split}
\end{equation}

The effect of the all-line shift on the subamplitudes is that all angle brackets are $O(1)$ and all square brackets are $O(z)$. The only exception is when the subamplitude $J$ has a total of 3 legs,
 namely
$\tilde{\ell}$, $\tilde{P}_I$ and, say, $m$. In that case, $| \mathring{P}_I \> \propto|m\>$ and $| \hat{\tilde{\ell}} \> \propto|m\>$. Hence $\<m \hat{\tilde{P}}_{I}\> \sim \<m \hat{\tilde{\ell}}\> \sim \<\hat{\tilde{\ell}} \hat{\tilde{P}}_{I}\>\sim O(z^{-1})$ and not $O(1)$ as all other angle brackets. However, this plays no role since the kinematics of the BCFW shift ensures that the 3-point vertex $J$ must be anti-MHV, and hence only depends on square brackets (if it were MHV in would vanish).

\emph{We conclude  that, at large $z$, an all-line square spinor shift
of the external lines acts to leading order as an all-line square spinor shift
on each of the two subamplitudes of any diagram in the BCFW expansion.}

\subsubsection*{Choice of parameters $c_i$}
In the above analysis we have implicitly assumed that the choice of complex parameters $c_i$ is \emph{generic}, \ie that the coefficients do not satisfy any accidental relations which affect the large $z$ behavior. For example, we assumed that $c_i\neq 0$ for all $i$, so that the shift is a genuine all-line shift. We also assumed that for all
subsets $\a$ of the external momenta, the $c_i$ satisfy
\begin{equation}\label{inequality}
    \sum_{i\in \a}  c_i\, \<\,i\,|  \neq 0\,.
\end{equation}
Then the internal line always shifts and the $O(z)$ terms in~(\ref{shiftedPI2}) and~(\ref{shiftedPI}) are indeed nonvanishing. Furthermore, we assumed that the subamplitudes are, to leading order, also subject to a generic all-line shift. In particular, this requires that
\begin{itemize}
  \item the coefficient $\tilde{c}_1$ is non-vanishing,
  \item all square brackets go as $[\,\hat{\tilde{i}}\,\hat{\tilde{j}}\,]\sim O(z)$ under the secondary shift, and
  \item the relation corresponding to~(\ref{inequality}) is also satisfied for the square spinor
  shift that acts on the subamplitudes $I$ and $J$.\footnote{Note that the relation~(\ref{inequality}) for the full amplitude does not immediately imply the corresponding relation for the subamplitudes, because the latter are evaluated at momenta $\tilde{p}_i$ shifted under the primary shift.}
\end{itemize}
We explicitly verified that all necessary inequalities are satisfied for a generic choice of the $c_i$. We thus only need to exclude a set of measure zero in the parameter space of the $c_i$.


\end{document}